\documentclass[english, reprint, aps, prb, twocolumn, superscriptaddress, floatfix]{revtex4-1}
   
\usepackage{amssymb}
\usepackage{amsmath}
\usepackage{float}
\usepackage{bbm}
\usepackage{graphicx}
\usepackage{verbatim}

\begin{document}

\title{On the validity of microscopic calculations of double-quantum-dot spin qubits based on Fock-Darwin states}

\author{Guo Xuan Chan}
\affiliation{Department of Physics and Materials Science, City University of Hong Kong, Tat Chee Avenue, Kowloon, Hong Kong SAR, China}
\author{Xin Wang}
\affiliation{Department of Physics and Materials Science, City University of Hong Kong, Tat Chee Avenue, Kowloon, Hong Kong SAR, China}
\affiliation{City University of Hong Kong Shenzhen Research Institute, Shenzhen, Guangdong 518057, China}

\begin{abstract}
We consider two typical approximations that are used in the microscopic calculations of double-quantum dot spin qubits, namely, the Heitler-London (HL) and the Hund-Mulliken (HM) approximations, which use linear combinations of Fock-Darwin states to approximate the two-electron states under the double-well confinement potential. We compared these results to a case in which the solution to a one-dimensional Schr\"odinger equation was exactly known and found that typical microscopic calculations based on Fock-Darwin states substantially underestimate the value of the exchange interaction, which is the key parameter that controls the quantum dot spin qubits. This underestimation originates from the lack of tunneling of Fock-Darwin states, which is accurate only in the case with a single potential well. Our results suggest that the accuracies of the current two-dimensional molecular-orbit-theoretical calculations based on Fock-Darwin states should be revisited since underestimation could only deteriorate in dimensions that are higher than one.
\end{abstract}

\maketitle

\section{Introduction}
Quantum computing has garnered much research interest owing to its potential in solving problems that are considered overwhelming for classical computers, such as factoring an integer in an efficient manner using Shor's algorithm\cite{shor1999polynomial} and high speed search in an unstructured search space using Grover's algorithm\cite{nielsen2002quantum}. Encoding of a qubit, which is the fundamental unit in quantum computation, has been proposed for a number of physical systems. In particular, spin qubits encoded in semiconductor quantum dots have been thought to be among the best candidates owing to their long coherence time, relatively high control fidelities\cite{petta2005coherent,koppens2008spin,barthel2012relaxation,maune2012coherent,pla2013high,muhonen2014storing,kim2014quantum,qi2017effects}, and their potential to be scaled up\cite{taylor2005fault}. With respect to the host material, namely, semiconductor quantum dots\cite{han2017radio,chen2017enhanced}, many qubit architectures have been proposed, including the single-spin qubit encoded in the spin states (spin-up or spin-down) of an electron\cite{muhonen2014storing,loss1998quantum}, the singlet-triplet (ST$_0$) qubit using singlet and triplet states of two-electron states as the computational basis\cite{petta2005coherent}, and the exchange-only qubit\cite{divincenzo2000universal,laird2010coherent} as well as its alternative version, the resonance-exchange qubit\cite{medford2013quantum}, can be realized by particular three-electron states in triple quantum-dot and hybrid qubits \cite{shi2012fast,kim2014quantum,PhysRevLett.116.086801,PhysRevB.95.035408}. Conventionally, the microscopic theoretical studies of double quantum dots are established by utilizing molecular orbital theory \cite{burkard1999coupled,hu2000hilbert,he2005singlet,saraiva2007reliability,li2010exchange,yang2011low,nielsen2012configuration,mehl2014inverted,calderon2015directly}. By adopting Fock-Darwin (F-D) states, the calculations based on the configuration interaction method are performed conveniently, and various qubit features have been derived, such as the exchange interaction and energy spectra of the ST$_0$ qubit along with their dependencies on external factors such as electric and magnetic field strength as well as their responses to environmental noises\cite{nielsen2010implications,raith2011theory,barnes2011screening,bakker2015validity}.
Several approaches have been proposed to investigate the exchange interaction in the ST$_0$ qubit. Among them, the simplest methods that are sufficient for capturing the qualitative information of the exchange interaction of the ST$_0$ qubit are the Heilter-London (HL) approximation and the Hund-Mulliken (HM) approximation, which is more accurate\cite{li2010exchange,burkard1999coupled}. In addition to the HL and HM approximations, a more complete calculation methodology, namely, the full configuration interaction calculation, was adopted to provide a precise result \cite{nielsen2010implications,nielsen2012configuration}. It was also shown that the characterization of valley-orbit coupling is crucial for complementing the study of energy splitting in the Si quantum dot qubit \cite{culcer2013dephasing,jiang2013coulomb,zimmerman2017valley}. In this paper, we propose an alternative approach to the above-mentioned approaches toward tackling the two levels of approximation that arise when HL or HM approximation is adopted in the study of the exchange interaction of the ST$_0$ qubit, which may lead to certain degrees of deviation and errors regarding  the calculation results. The first level of approximation is made when multiple-electron wave functions in double or multiple quantum dots are constructed by single-well single-electron Schr\"odinger equation solutions, namely, the Fock-Darwin (F-D) states. Secondly, with the assumption of negligible mixing between excited states and the ground state, only the ground-state wave function is considered in the construction of multiple-electron wave functions. In this paper, we attempted to relax the first level of approximation by utilizing the solutions of exactly solvable single-particle double-well Schr\"odinger equation, presented by Caticha (1995)\cite{caticha1995construction}, as a replacement of Gaussian function (F-D states) to construct two-electron wave functions in the double quantum dot for calculating the exchange interaction of the ST$_0$ qubit. Since the exactly solvable double-well solutions allow electron tunnelling between two dots, which is largely underestimated by Fock-Darwin states, we would expect a more significant absolute value of exchange interaction.

This paper is organized as follows. In Sec.\ref{sec:model}, the model and methods used to calculate the exchange interaction are presented. The calculation results are presented in Sec.\ref{sec:results} including the exchange interaction as a function of interdot distance (Sec. \ref{subsec:distance}) and as a function of well depth (Sec. \ref{subsec:depth}). Finally, we conclude our paper in Sec. \ref{sec:conclusions}.

\section{Model and Methods} 
	\label{sec:model}
\subsection{Conventional method of describing electron states}
The potential confinement experienced by electrons in the double quantum dot (DQD) device is well described or approximated by the two-dimensional potential function owing to the high degree of confinement in \(\hat{z} \) direction\cite{li2010exchange}. However, the availability of exactly solvable double-well potential solutions is limited to one-dimensional solutions\cite{chen2012heun,xie2015analytical,xie2012new,jelic2012double,munoz2014exactly,caticha1995construction} and has not yet been developed for two-dimensional solutions. Therefore, in this study, we will adopt the solution of the one-dimensional double-well potential Schr\"odinger equation presented by Caticha\cite{caticha1995construction} to investigate the exchange coupling in $\mathrm{ST_0}$ qubit architecture.

The DQD device fabricated by either Si/SiO$_2$ or GaAs allows us to discuss the lowest-energy state by disregarding the mixing of the higher energy states into the ground state\cite{burkard1999coupled}. In particular, for the DQD grown at the Si/SiO$_2$ interface, due to the \(\hat{z} \) direction confinement that leads to the splitting of the six-fold degeneracy in bulk Si along with the further splitting of the ground energy state due to interface potential \cite{li2010exchange}, the exchange coupling and other energy features of the qubit architecture can be solely investigated using the lowest-energy state.
Although the Fock-Darwin states [Eq. \eqref{eq:FDSt}] are not the exact solutions to the double-well potential function Schr\"odinger equation, they are conventionally adopted to describe the single electron state in DQD\cite{burkard1999coupled,li2010exchange}, under the assumption that near the bottom of the well (electron occupying the ground energy state), the potential function can be approximated to be bi-quadratic [Eq. \eqref{eq:BQPo}] \cite{burkard1999coupled,li2010exchange}. Nevertheless, the exchange coupling and other qubit features calculated with respect to Fock-Darwin states are subject to deviation and errors since the Fock-Darwin states being merely an approximation while describing electron states in DQD. This approach fails to address two basic issues that emerge as a result of the double-well potential function. First, the Fock-Darwin state can only solve electron states in a single well (quadratic potential well). Second, as a result of the first issue, the tunneling effect between the two wells is substantially truncated for a single electron state, which means that the electron occupying one well has a comparatively diminished tunnelling probability. The previous mentioned issues are addressed when exactly solvable double-well potential solutions are adopted for constructing the electron states in the DQD. Therefore, we employ the exact solutions of the double-well potential Schr\"odinger equation to investigate the validity of the results obtained using the Fock-Darwin states.
\begin{equation} \label{eq:BQPo}
V^\mathrm{BQ}(x)= \frac{1}{2} m \omega_0^2 (\mathrm{Min}[(x-d)^2,(x+d)^2])
\end{equation}
\begin{equation} \label{eq:FDSt}
\psi^\mathrm{FD}_{L,R}(x)= \frac{1}{\sqrt{a}\pi^{1/4}} \exp \left[-\frac{(x \pm d)^2}{2a^2}\right]
\end{equation}

\subsection{Calculation of exchange interaction}
The fundamental physical quantity for quantum computing in the ST$_0$ qubit architecture is the energy splitting between  the unpolarized triplet state, $|T_0\rangle$, and 
the singlet state, $|S\rangle$. Energy splitting is referred to as the exchange interaction, which is $J\equiv E_{T_0} - E_S$. If there are two electrons in the DQD and the interaction between them is limited to the Coulomb interaction \cite{burkard1999coupled,li2010exchange}, there will only be four spin eigenstates in a uniform magnetic field, namely, 
a singlet state (
$|S\rangle = \frac{1}{2}|\uparrow\downarrow\rangle - |\downarrow\uparrow\rangle$ 
with a total spin of $ S = 0$) and three triplet states ($|T_{0,+,-}\rangle = \frac{1}{2}|\uparrow\downarrow\rangle + |\downarrow\uparrow\rangle, |\uparrow\uparrow\rangle,|\downarrow\downarrow\rangle $ 
With a total spin of $ S=1 $). With respect to the four spin eigenstates outlined above, the Hamiltonian can be written as follows: 
\begin{equation}
 \hat{H} = \sum\limits_{i=1,2}^{} \hat{h}_i + \frac{e^2}{\kappa |\mathbf{r}_{12}|} = \sum\limits_{i=1,2}^{} \hat{h}_i + \hat{C}
\end{equation}
where $ i = 1,2 $ corresponds to the two electrons in DQD, $ \kappa $ is the effective dielectric constant $ \kappa = (\epsilon_{Si} + \epsilon_{SiO_2})/2 $, $ e $ is the electron's electrical charge and $ |\mathbf{r}_{12}| = |\mathbf{r}_1 - \mathbf{r}_2| $ is the absolute separation value between two electrons. $ \hat{h}_i $ is the single-particle Hamiltonian where:
\begin{equation}
\hat{h}_i = \hat{T}_i + V(\mathbf{r}_i) + e E x_i + g_\mathrm{eff} \mu_B B S_{iz}
\end{equation}
\begin{equation}
\hat{T}_i = \frac{1}{2 m} \left[\mathbf{p}_i - \frac{e}{c} \mathbf{A} (\mathbf{r}_i) \right]^2
\end{equation}
$ m $ is the effective mass in the traverse direction ($ \hat{z} $ and $-\hat{z}$ direction) ($ m = 0.191 m_e$ for Si/SiO $_2$ DQD and $ m = 0.067 m_e$ for GaAs DQD). $ \mathbf{A}(\mathbf{r}_i) $ is the vector potential of the magnetic field in the $ \hat{z}$ direction, $ \mathbf{A} = B(-y,x,0)/2 $. However, the vector potential will not be considered in our discussion since the exactly solvable solutions are limited to one-dimensional solutions. $ E $ is the electric field applied along the $ \hat{x}$ direction to displace the potential well in the $ \hat{z}$ direction. The term $ g_\mathrm{eff} \mu_B B S_{iz} $ is the Zeeman energy of the electrons. $ V(\mathbf{r}_i) $ is the potential confinement experienced by the electrons in the DQD. Conventionally, the bi-quadratic potential function [Eq. \eqref{eq:BQPo}] is employed as an approximation, as discussed in the previous section. In this study, we made a comparison between the conventional calculation method and the exactly solvable double-well potential [Eq. \eqref{eq:CaPo}], which will be introduced in a later section.

The calculation of the exchange interaction can be performed by two approaches, namely, Heitler-London (HL) and Hund-Mulliken (HM), which will be described in detail further in the paper.
\subsection{Heitler-London}
The HL approach can be considered as the simplest estimation of the exchange interaction as only single-occupied states ($|S(1,1)\rangle$, singlet and $|T_0(1,1)\rangle$, triplet) are considered under the assumption that doubly occupied singlet states, namely, $|S(2,0)\rangle$ and $|S(0,2)\rangle$, have much higher energy, such that the mixing between $|S(1,1)\rangle$ with $|S(2,0)\rangle$ or $|S(0,2)\rangle$ does not occur. On the basis of the above assumption, the electron states in the DQD can be constructed from single electron ground-state orbital wave functions into two-electron orbital state wave functions, as an artificial hydrogen molecule. Hence, the spatial part of the electron states is represented as follows:
\begin{equation}
|S/T_0\rangle = \frac{| \psi_L(1)\psi_R(2) \rangle \pm |\psi_R(1)\psi_R(2) \rangle}{\sqrt{2(1\pm l^2)}}
\end{equation}
$l$ is the overlap between the left and right dot single electron state $l=\langle\psi_L|\psi_R\rangle$
The exchange interaction, which is the energy splitting between the singlet state, $|S\rangle$, and unpolarized triplet state, $|T_0\rangle$, is represented as follows:
\begin{equation}
J_\mathrm{HL} = \langle T_0 | \hat{H} | T_0 \rangle - \langle S | \hat{H} | S \rangle
\end{equation}
which can be rewritten in a simpler form as:
\begin{equation} \label{JHL}
J_\mathrm{HL} = \frac{2 l^2}{1 - l^4} \left(W_v + D_0 - \frac{1}{l^2} E_0 \right) 
\end{equation}
The explicit form of each element in [Eq. \eqref{JHL}] is:
\begin{equation} \label{eq:Wv}
W_v = \langle \psi_L(1)\psi_R(2) | \hat{v} | \psi_L(1)\psi_R(2) - \frac{1}{l^2} \psi_R(2)\psi_L(1) \rangle
\end{equation}
where $\hat{v} = \hat{h_1} + \hat{h_2}$, or more explicitly as $\hat{v} = V(1)+V(2)-V_L(1)-V_R(2)$ when Fock-Darwin states are adopted.
\begin{equation} \label{eq:D0}
D_0 = \langle \psi_L(1)\psi_R(2) | \hat{C} | \psi_L(1) \psi_R(2) \rangle
\end{equation}
\begin{equation}
E_0 = \langle \psi_L(1)\psi_R(2) | \hat{C} | \psi_R(1) \psi_L(2) \rangle
\end{equation}
$W_v$ is the kinetic energy gain of the singlet state, $D_0$ is the contribution of the Coulomb interaction, and $E_0 $ is the exchange Coulomb interaction. It can be seen that in any case in which the exact solutions and their corresponding potential functions are adopted, the $W_v$ term has zero value since all the terms in its explicit form cancel each other.

\subsection{Hund-Mulliken}
Nevetheless, the HL approach fails to consider the doubly occupied electron states that are necessary not only for the initialization and measurement of qubit states but also for the capacitive coupling required by the two-qubit manipulations. This problem was addressed when adopting the Hund-Mulliken (HM) approach, which takes two additional doubly occupied states, namely, $|S(2,0)\rangle$ and $|S(0,2)\rangle$, into the Hamiltonian $\hat{H}$, and provides a $4\times4$ matrix. The HM approach, with its more rigorous estimation, has been proven to provide a more accurate estimation of the exchange interaction $J$ \cite{burkard1999coupled,li2010exchange}.

First, the single-electron state wave function has to be orthonormalized into $\Phi_{L/R} = (\psi_{L/R} - g \psi_{R/L}) / \sqrt{1-2 l g-g^2}$ where $g = (1 - \sqrt{1 - l^2})/l$. With the orthonormalized single-electron state wave functions, two-electron states in DQD, which include the two doubly occupied singlets $|S(2,0)\rangle$ and $|S(0,2)\rangle$, the (1,1) singlet $|S(1,1)\rangle$, and the (1,1) triplet $|T(1,1)\rangle$, can be constructed.
\begin{equation}
|S(2,0)/(0,2)\rangle = \Psi_{L/R}^d(\mathbf{r}_{1},\mathbf{r}_{2}) = \Phi_{L/R}(\mathbf{r}_1)\Phi_{L/R}(\mathbf{r}_2)
\end{equation}
\begin{equation}
\begin{split}
|S/T(1,1)\rangle = \Psi^{(1,1)}_{S/T}(\mathbf{r}_{1},\mathbf{r}_{2})& = \\
\frac{1}{\sqrt{2}} [\Phi_{L}(\mathbf{r}_1)\Phi_{R}(\mathbf{r}_2) &\pm \Phi_{L}(\mathbf{r}_2)\Phi_{R}(\mathbf{r}_1)]
\end{split}
\end{equation}
With respect to the four bases shown above, obtaining the eigenstates and the corresponding eigenvalues is a four-dimensional problem. In other words, the Hamiltonian is a $4\times4$ matrix
\begin{equation}
H =
\begin{pmatrix}
 U - \epsilon & X & \sqrt{2}t & 0\\ 
 X & U + \epsilon & \sqrt{2}t & 0 \\
 \sqrt{2}t & \sqrt{2}t & V_S & 0 \\
 0 & 0 & 0 & V_T
\end{pmatrix}\label{eq:4 matrix}
\end{equation}
where:
\begin{equation}
\epsilon = \epsilon_R - \epsilon_L
\end{equation}
\begin{equation}
\begin{split}
\epsilon_{L/R} = \langle \Phi_{L/R}(\mathbf{r}_1) | \hat{h}_1 | \Phi_{L/R}(\mathbf{r}_1) \rangle \\
= \langle \Phi_{L/R}(\mathbf{r}_2) | \hat{h}_2 | \Phi_{L/R}(\mathbf{r}_2) \rangle 
\end{split}
\end{equation}
\begin{equation}
X = \langle \Psi^d_{L/R}(\mathbf{r}_1,\mathbf{r}_2) | \hat{C}_{12} | \Psi^d_{R/L} (\mathbf{r}_1,\mathbf{r}_2)\rangle
\end{equation}
\begin{equation} 
U = \langle \Psi^d_{L/R} (\mathbf{r_1},\mathbf{r}_2) | \hat{C}_{12} | \Psi^d_{L/R} (\mathbf{r}_1,\mathbf{r}_2) \rangle
\end{equation}
\begin{equation}
V_S = \langle \Psi_S^{(1,1)}(\mathbf{r}_1,\mathbf{r}_2) | \hat{C}_{12} | \Psi_S^{(1,1)} (\mathbf{r}_1,\mathbf{r}_2) \rangle
\end{equation}
\begin{equation}
V_T = \langle \Psi_T^{(1,1)}(\mathbf{r}_1,\mathbf{r}_2) | \hat{C}_{12} | \Psi_T^{(1,1)} (\mathbf{r}_1,\mathbf{r}_2) \rangle
\end{equation}
\begin{equation}
\begin{split}
t = t' + w = \langle \Phi_{L/R}(\mathbf{r}_{1/2}) | \hat{h}_{1/2} | \Phi_{R/L}(\mathbf{r}_{1/2}) \rangle\\
+ \frac{1}{\sqrt{2}}\langle\Psi^d_{L/R}(\mathbf{r}_1,\mathbf{r}_2)| \hat{C} | \Psi_S^{(1,1)}(\mathbf{r}_1,\mathbf{r}_2) \rangle
\end{split}\end{equation}
Here, $\epsilon_L$ and $\epsilon_R$ are single-particle energies of the electron occupying the left and right dot, respectively. $X$ is the interdot Coulomb interaction (Coulomb interaction between electrons in different dots). $U$ is the intradot Coulomb interaction (Coulomb interaction between electrons in the same dot), $t$ is the modified interdot tunneling interaction, and $t$' is the bare tunneling (electron hopping) term, while $w$ is the Coulomb term.
By obtaining the eigenvalues and eigenvectors of the $4\times4$ Hamiltonian matrix (Equation~\eqref{eq:4 matrix}),
we can obtain the energy diagram of the two-dot system. Each obtained eigenvalue 
corresponds to the energy value of the state eigenvector. The state vector can be a pure basis
state or a mixed state with different bases having different probabilities. By calculating the
numerical values for each matrix element, which are different for different conditions (e.g.,
different potential well separation, different detuning strength), the energy of each electron
state can be known; thus, the exchange interaction between the singlet state $|S\rangle$ and triplet state $|T_0\rangle$ can be obtained by subtracting the energy value of the triplet state by the singlet state.

\begin{figure}
  \centering
  \includegraphics[width=\columnwidth]{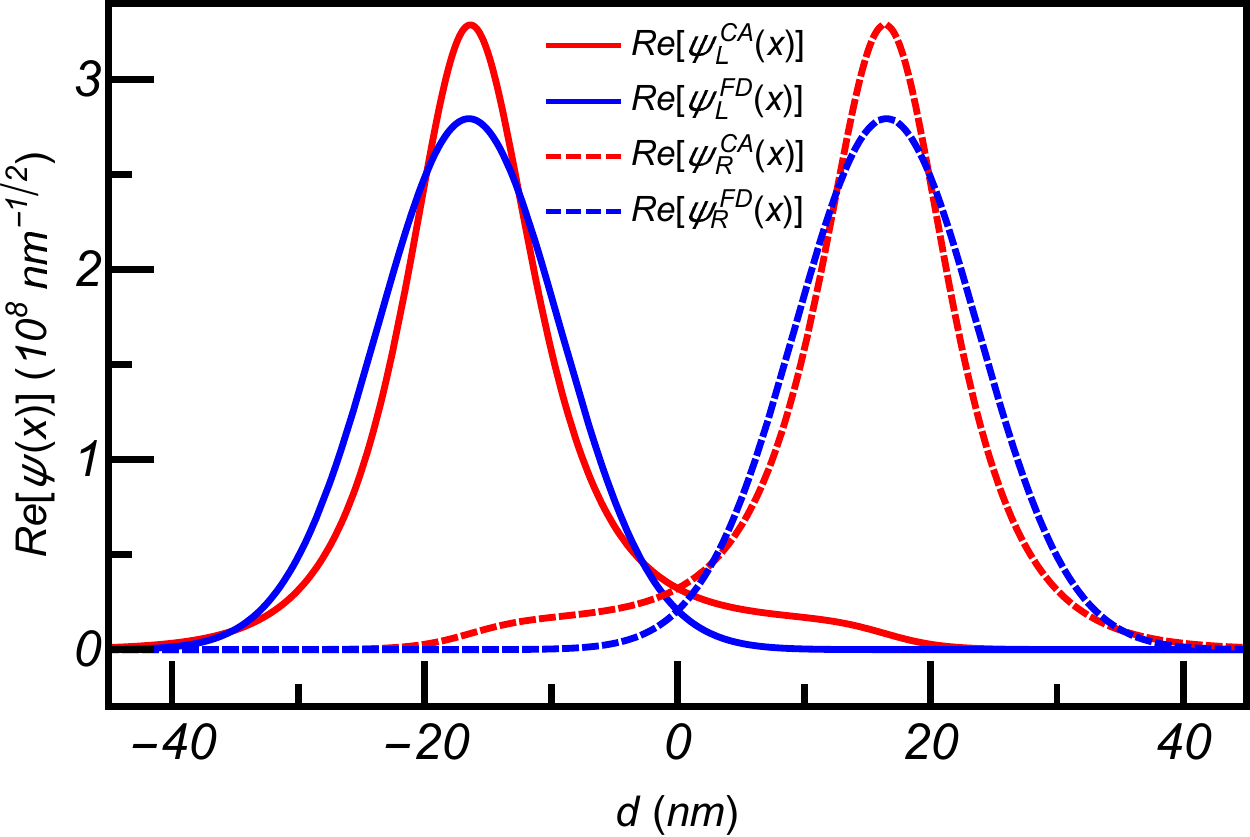}
  \caption{Comparison of Caticha and real part of Fock-Darwin wave functions. Caticha wave functions ($ \psi^{\mathrm{CA}}_L $ ($ \psi^{\mathrm{CA}}_R $)) [Eq.~\eqref{eq:CaSt}] are indicated by red solid (dashed) lines for the left (right) dot, respectively, while the real part of the Fock-Darwin wave functions ($ \psi^{\mathrm{FD}}_L $ ($ \psi^{\mathrm{FD}}_R $)) (Eq \ref{eq:FDSt}) is indicated by blue solid (dashed) lines.}
  \label{fig:CaWavefunction}
\end{figure}

\subsection{Exactly solvable double-well potential}
	\label{subsec:exact}
The exactly solvable double-well potential Schr\"odinger equation has been solved by a number of mathematicians using different approaches \cite{chen2012heun,xie2015analytical,xie2012new,jelic2012double,munoz2014exactly,caticha1995construction}. In this study, the exchange interaction was calculated by adopting the solutions presented by Caticha \cite{caticha1995construction}. The reason for choosing Caticha's solution among other solutions was mainly because of the feasibility of manipulating the wells parameters, namely wells separation and well depth, as well as because of the simplicity of the potential function. However, there has not been any exact solution to the two-dimensional double-well potential Schr\"odinger equation; hence, our discussion and calculation results were derived from the one-dimensional solutions provided by Caticha\cite{caticha1995construction}.

\begin{figure}
  \centering
  \includegraphics[width=\columnwidth]{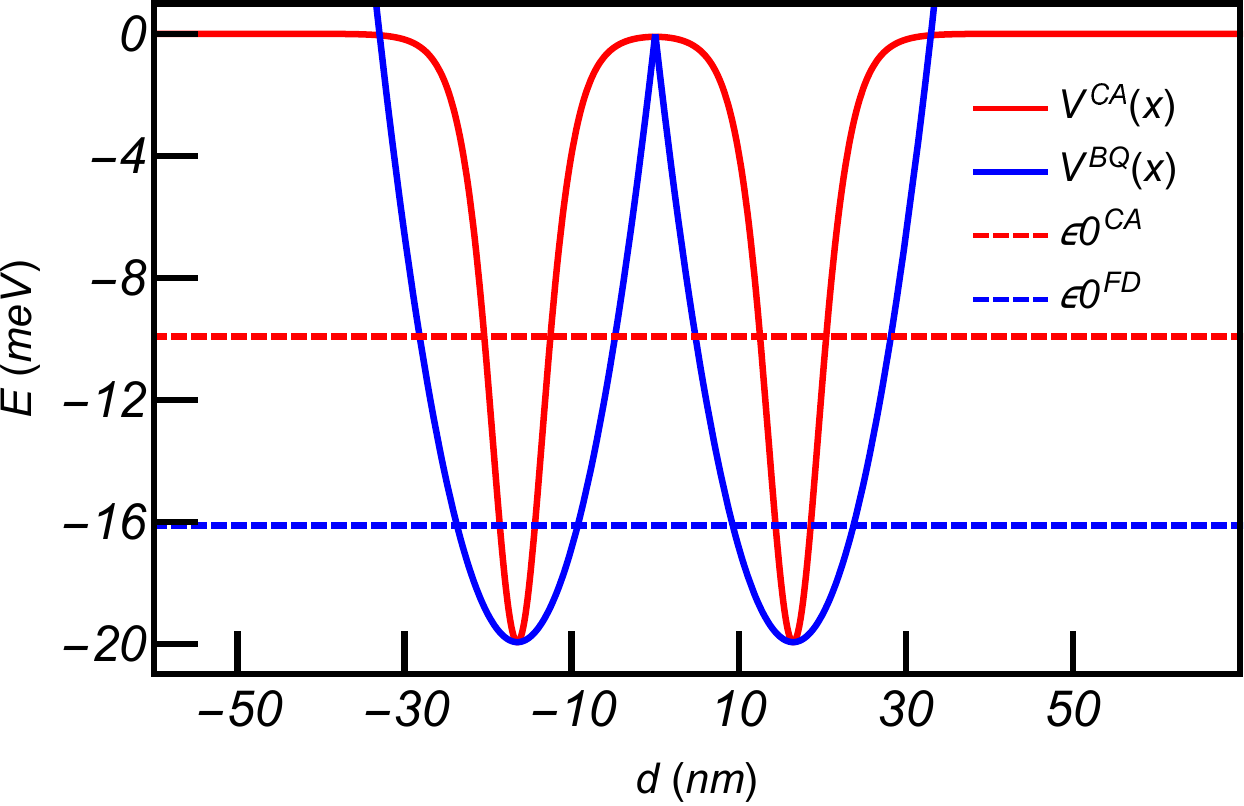}
  \caption{Comparison of Caticha and bi-quadratic confinement potential. The Caticha potential (Eq.\ref{eq:CaPo}) is indicated by the red solid line, while the bi-quadratic potential (Eq.\ref{eq:BQPo}) is indicated by the blue solid line. The horizontal red (blue) dashed lines indicate the ground state energies for the Caticha(bi-quadratic) potential. Parameters: $a = 2.2361 \times 10^8$, $b = 2.2305 \times 10^8$, $\hbar \omega_0= 7.658 meV $.}
  \label{fig:CaPotential}
\end{figure}

The ground-state wave function that can exactly solve the double-well potential Schr\"odinger Equation, presented by Caticha\cite{caticha1995construction}, is:
\begin{equation} \label{eq:CaSt}
\begin{split}
\psi &^\mathrm{CA}_{L/R}(x)=\\
\frac{1}{c} &\frac{2a \, e^{\pm ax} (1+e^{
\pm bx})}{(a+b)(e^{\pm 2ax}+e^{\pm 2bx})+(a-b)(e^{\pm 2(a+b)x}+1)}
\end{split}\end{equation}
with the potential function as: 
\begin{equation} \label{eq:CaPo}
V ^\mathrm{CA}(x)=-2 \frac{(a^2-b^2)[a^2 \cosh^2(bx)+b^2 \sinh^2(ax)]}{[a \, \cosh(ax) \cosh(bx)-b \, \sinh(ax) \sinh(bx)]^2}
\end{equation}
\textit{a} and \textit{b} are the parameters that control the features of the potential well, namely, the well separation (also called interdot distance) \textit{d} and well depth $\xi$. The well features were controlled by the absolute value of \textit{a} and \textit{b} and the relationship between them. First, the value of \textit{a} had to be positive and larger than \textit{b} (\(a > b \)), such that the potential function would provide a double-well characteristic with a maximum value of 0, which corresponds to the energy of a free particle. If \(a < b \), the potential function showed double barrier characteristics instead, with a minimum value of 0. Second, the well separation was controlled by the difference between \textit{a} and \textit{b}, and the larger the difference was, the closer the well separation became. Third, the absolute magnitude of the value \textit{a} and \textit{b} determined the depth of the double wells, and the larger the absolute values of \textit{a} and \textit{b} were, the deeper the wells were. The explicit values of the well separation and well depth could not be determined analytically, instead, they could only be calculated numerically. In addition, the barrier height between the two wells was indirectly determined by the interdot distance and well depth; the deeper the well depth or the larger the interdot distance was, the higher was the potential barrier. The probability amplitudes of the wave functions shown in Fig. \ref{fig:CaWavefunction} (Red line), with the values of $a$ and $b$ as $ 2.2361 \times10^8 $ and $2.2305 \times 10^8$, respectively, could solve the Schr\"odinger equation with the double-well potential function shown in Fig. \ref{fig:CaPotential} (Red line). As a comparison, the blue lines in Fig. \ref{fig:CaWavefunction} and Fig. \ref{fig:CaPotential}, which are the Fock-Darwin states and the bi-quadratic potential function respectively, were plotted.  

For convenience, Fock-Darwin states and the exactly solvable wave functions presented by Caticha \cite{caticha1995construction} will be referred to as $\psi^{\mathrm{FD}}$ and $\psi^{\mathrm{CA}}$, respectively. Fig. \ref{fig:CaWavefunction} shows that CA's probability density was more concentrated at the interdot region than FD's probability density. Accordingly, a strong interdot (long range) Coulomb interaction and high possibility of interdot hopping would be expected. In fact, a high probability density concentration at the interdot region was considered as an unusual CA feature for the following two reasons: the wave functions were expected to be symmetric since both the left and right wells were symmetric; it showed stronger confinement for the potential function of CA than the bi-quadratic potential (Fig. \ref{fig:CaPotential}) with the same barrier height and interdot distance. However, this feature can be explained by the existence of a second potential well right next to the first well, which allowed the particle in one well to tunnel through the potential barrier and appear in another well. Additionally, since the FD ground state can only solve a single potential well Schr\"odinger equation, it does not have an asymmetric feature such as the CA state. In our study, this special feature of the CA state was the primary motivation for using the exactly solvable double-well potential, namely, the CA states, instead of using the approximate solutions, namely, the FD states.
\subsection{Numerical Calculation}
Conventionally, integrations involving the Fock-Darwin states can sometimes have analytical results for both Coulomb and kinetic terms. However, owing to the high complexity of the exactly solvable wave functions presented by Caticha \cite{caticha1995construction}, analytical solutions are not possible; hence, numerical integration methods were adopted. The integration boundaries and step (small steps for incrementing the evaluation values of integral functions for each iteration) had to be chosen carefully, such that the boundaries were wide enough to cover all high contribution areas, and the step was small enough to avoid the discrepancy of over-evaluation or under-evaluation. Proper integration boundaries and steps were chosen after performing a number of trails by increasing the range of boundaries and decreasing the step size until there was no significant change to the integration result. Numerical integration was performed by following a series of steps in the following manner:\\
(i) Start from one of the boundaries (usually the smaller value)\\
(ii) Evaluate the integral function numerical value at that position\\
(iii) Increment the evaluation position by adding the magnitude of the step to the initial
position\\
(iv) Add the numeric value of the integral function at the new position to the previously
evaluated numeric value\\
(v) Repeat step (i) till (iv) and the numeric value of each position accumulates until the evaluation position reaches the other boundary (larger value).

\section{Results} 
	\label{sec:results}
\subsection{Exchange coupling as a function of interdot distance, \(d\)}
	\label{subsec:distance}
In this section, the exchange coupling was calculated as a function of interdot distance with a well depth of $10 meV$. At the same time, a comparison to the exchange coupling calculated based on FD states was made. During the calculation, the well depth and potential barrier of the bi-quadratic potential were made the same as the potential function of CA for the same interdot distance.

\begin{figure}[h]
\centering{
 \includegraphics[width=.9\columnwidth]{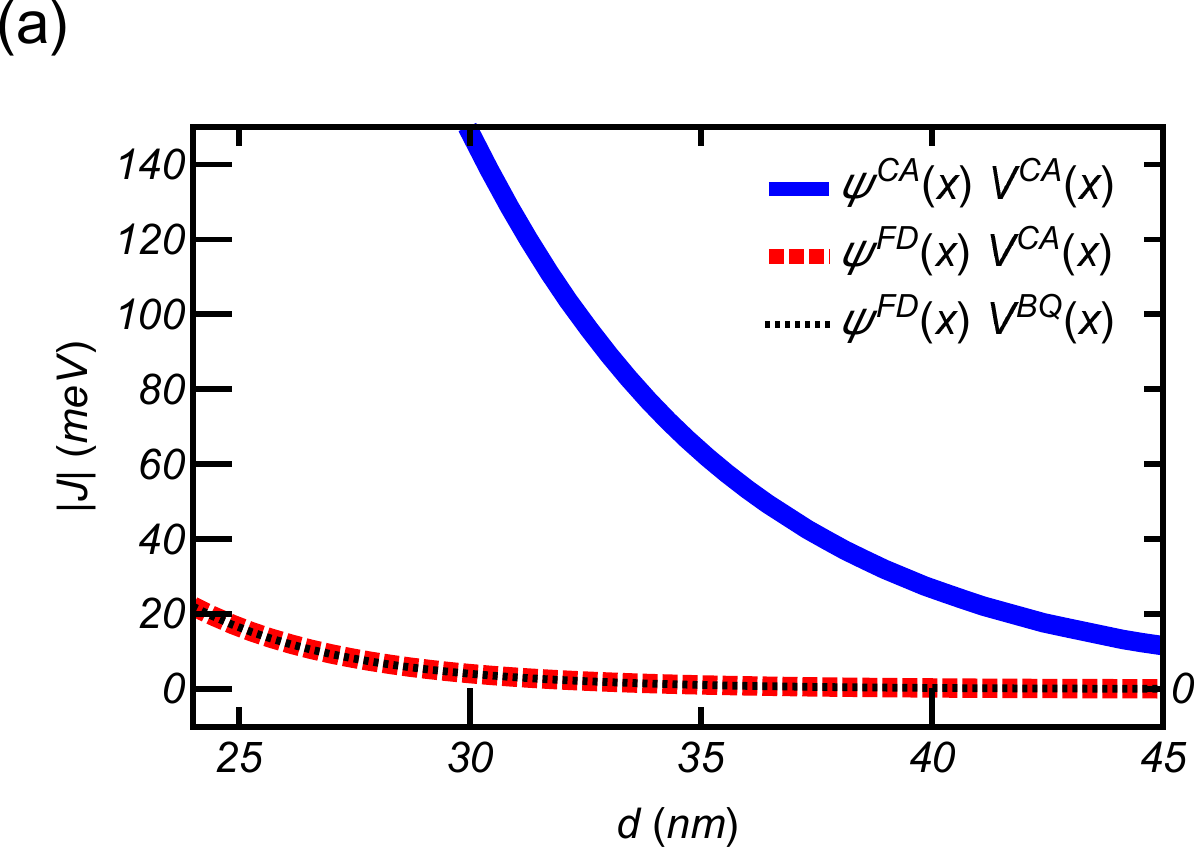}
 \hspace{1cm}
  \includegraphics[width=.9\columnwidth]{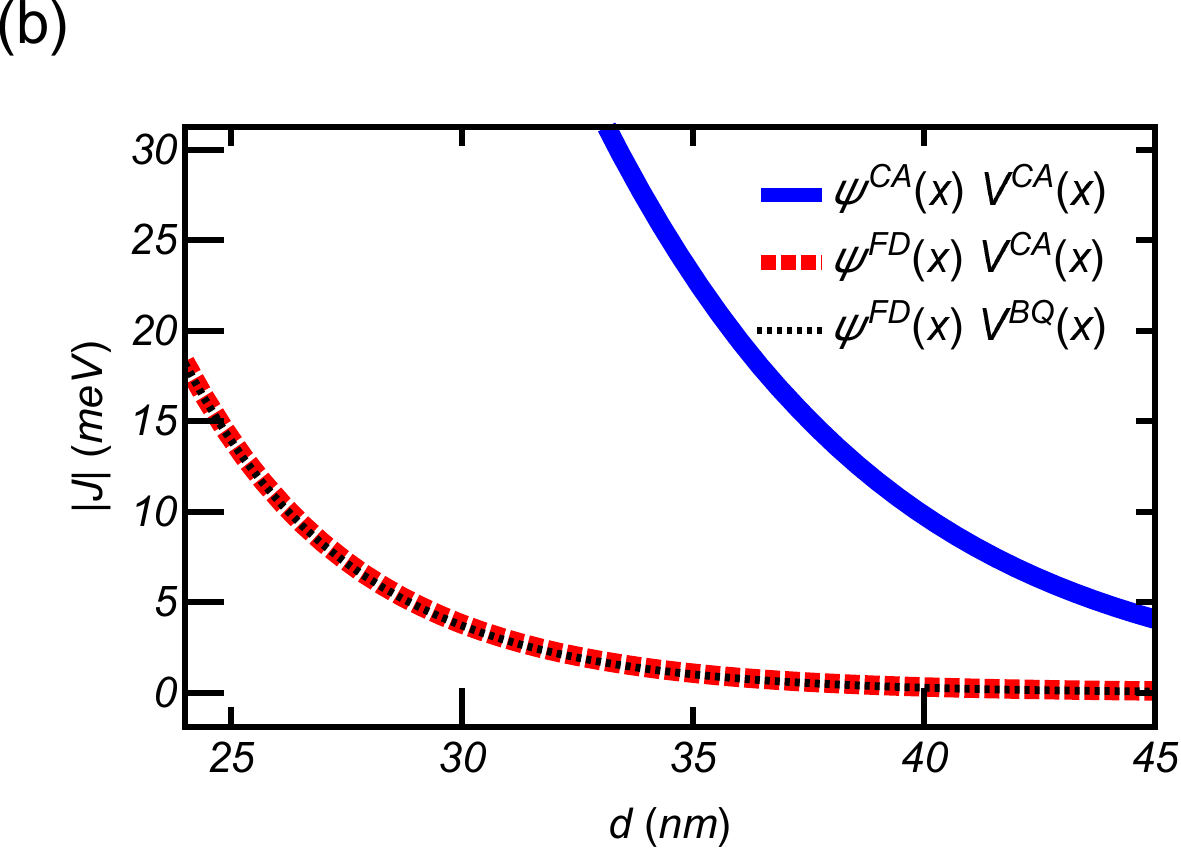}}
\caption{Calculated exchange interaction vs. interdot distance $d$ at well depth \textit{10 meV} by using (a) HL calculation and (b) HM calculation. The solid blue line corresponds to the exchange energy calculated by using Caticha's wave functions \(\psi ^\mathrm{CA}\) and Caticha's potential \(V ^\mathrm{CA}\). The dashed red line corresponds to the use of Fock-Darwin states \(\psi ^\mathrm{FD}\) along with Caticha's potential \(V ^\mathrm{CA}\) in the calculation. The dotted black line corresponds to the use of Fock-Darwin states \(\psi ^\mathrm{FD}\) along with the bi-quadratic potential \(V ^\mathrm{BQ}\).
}
\label{fig:interdot}
\end{figure}

The most prominent feature that can be observed in Fig. \ref{fig:interdot} is that the exchange interaction calculated by the exact solution proposed by Caticha \cite{caticha1995construction} was far greater than that calculated by FD states along with either the CA potential or BQ (harmonic) potential. This was due to the higher concentration of probability density at the interdot region, which led to stronger interdot (long range) Coulomb interaction and interdot hopping (cf. Fig.~\ref{fig:CaWavefunction}). Furthermore, Fig. \ref{fig:interdot} shows that the larger the interdot distance was, the weaker was the exchange interaction. This result can be explained by the degree of overlapping between the left and right dot wave functions ($\psi_L$ and $\psi_R$). The larger the interdot distance was, the less was the overlapping between the left and right dot, which led to weaker Coulomb interaction and interdot hopping; this resulted in a weak exchange interaction. Fig. \ref{fig:interdot} also shows that the exchange interaction predicted by the FD states  along with either the bi-quadratic or CA potential overlaps made no difference at all. This indicated that the exchange interaction that was correlated to the FD states was insensitive to the confinement strength exerted by the surrounding potential.

In addition, since the HL approach only concerned two states, namely, $|S(1,1)\rangle$ and $|T_0\rangle$ states, the exclusion of the $|S(2,0)\rangle$ and $|S(0,2)\rangle$ states may have led to a certain degree of error while calculating the exchange interaction. Therefore, the HM approach provided a more accurate result for the exchange interaction since the contribution of the $|S(2,0)\rangle$ and the $|S(0,2)\rangle$ states to the Hamiltonian $\hat{H}$. Fig. \ref{fig:interdot} was considered, which indicated that the exchange interaction predicted by the HL approach was larger by at least five times than that predicted by the HM approach. This suggested that the inclusion of the doubly occupied state in the HM approach had significant influence on the strength of the exchange interaction. Based on that, with its more robust approximation, HM provided a much more accurate estimation of exchange interaction than the HL model. Nevertheless, the exchange interaction provided the same qualitative result, namely, the exponential decay for the large interdot distance $d$, for both the HL and HM approaches.
However, we observed that there was an exponential increase in the exchange interaction when $d \lesssim 35 nm $ for both the HL and HM approximations, which was rather unphysical\cite{li2010exchange,burkard1999coupled}. A closer look to the difference between the HL and HM approximation result revealed that the above-mentioned exponential increase was less severe for the HM approximation, which indicated that the HM provided a more physical result than the HL approximation, particularly when the interdot distance was small. Moreover, it was observed that the obtained exchange interaction was in the range of $meV$ instead of $\mu eV$, which has been widely observed in previous studies\cite{li2010exchange,burkard1999coupled}. The inconsistent results can be attributed to the adoption of the one-dimensional wavefunction rather than the two-dimensional wavefunction while describing the single electron state in DQD using the proposed approach.

\begin{figure}
\centering{
 \includegraphics[width=.9\columnwidth]{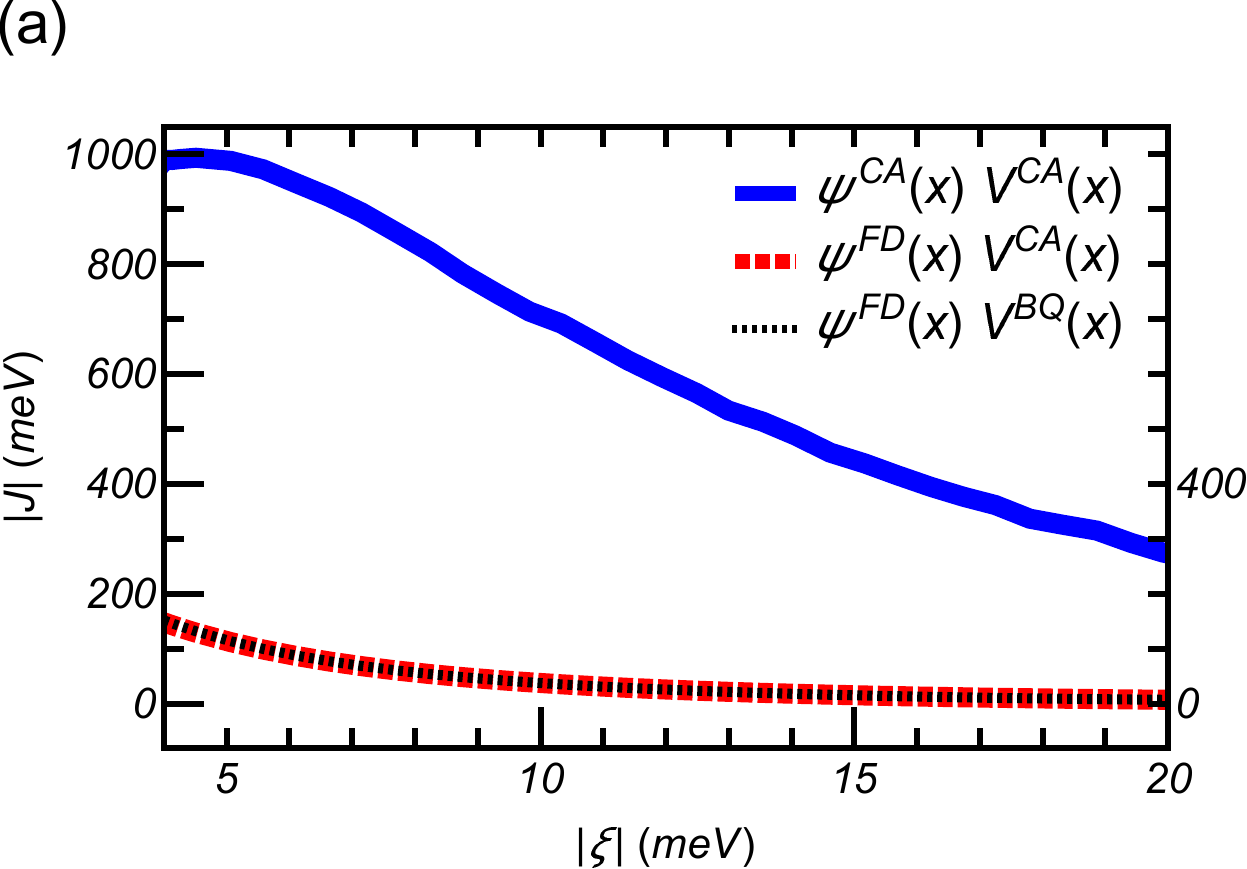}
 \hspace{1cm}
  \includegraphics[width=.9\columnwidth]{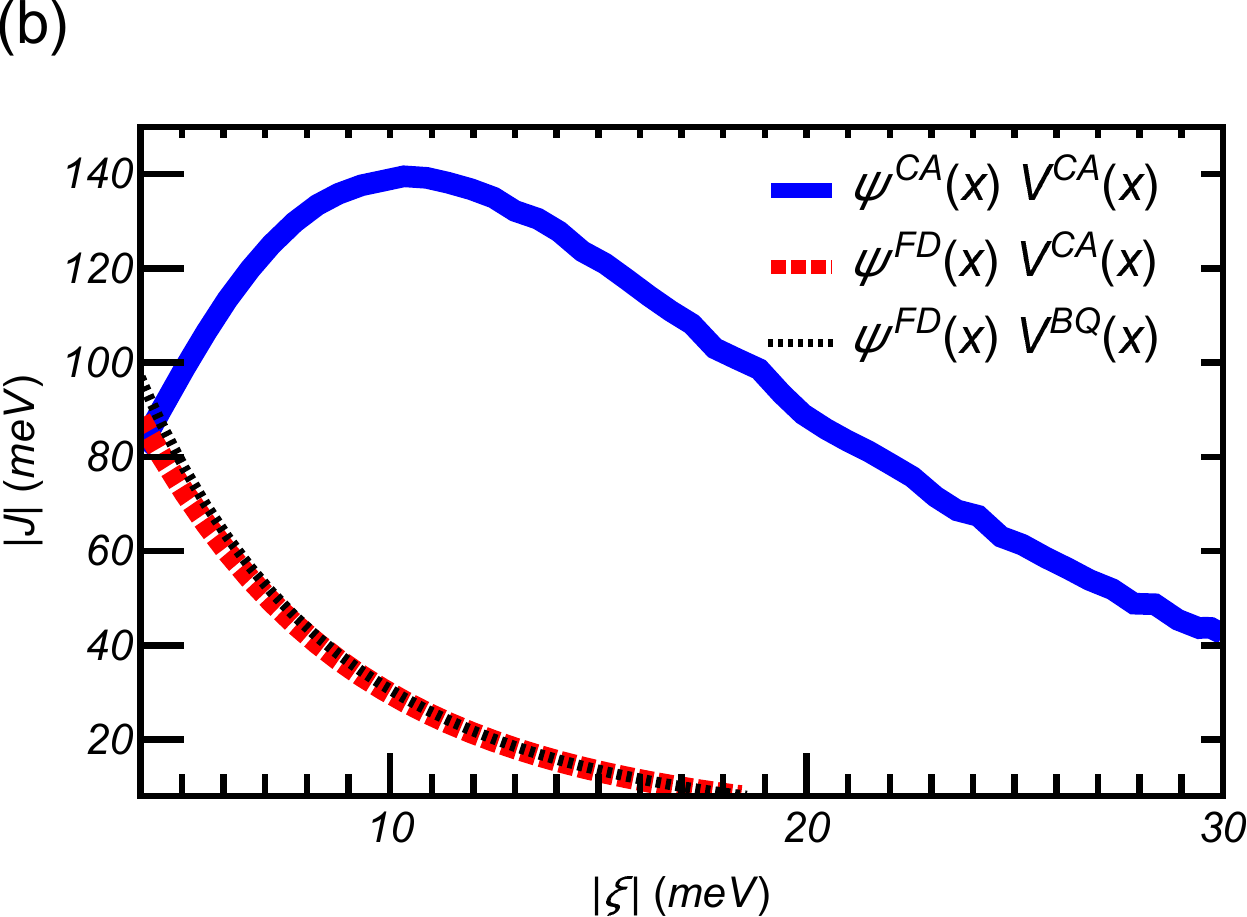}
  }
\caption{Calculated exchange energy vs. well depth \textit{\(|\xi| \)} at interdot distance \textit{20nm} by using (a) HL calculation and (b) HM calculation. The solid blue line corresponds to the exchange energy calculated using Caticha's wave functions \(\psi ^\mathrm{CA}\) and Caticha's potential \(V ^\mathrm{CA}\). The dashed red line corresponds to the use of Fock-Darwin states \(\psi^{FD}\) along with Caticha's potential \(V ^\mathrm{CA}\) in the calculation. The dotted black line corresponds to the use of Fock-Darwin states \(\psi ^\mathrm{FD}\) along with the bi-quadratic potential \(V ^\mathrm{BQ}\).}
\label{fig:depth}
\end{figure}

\begin{figure}
  \centering
  \includegraphics[width=\columnwidth]{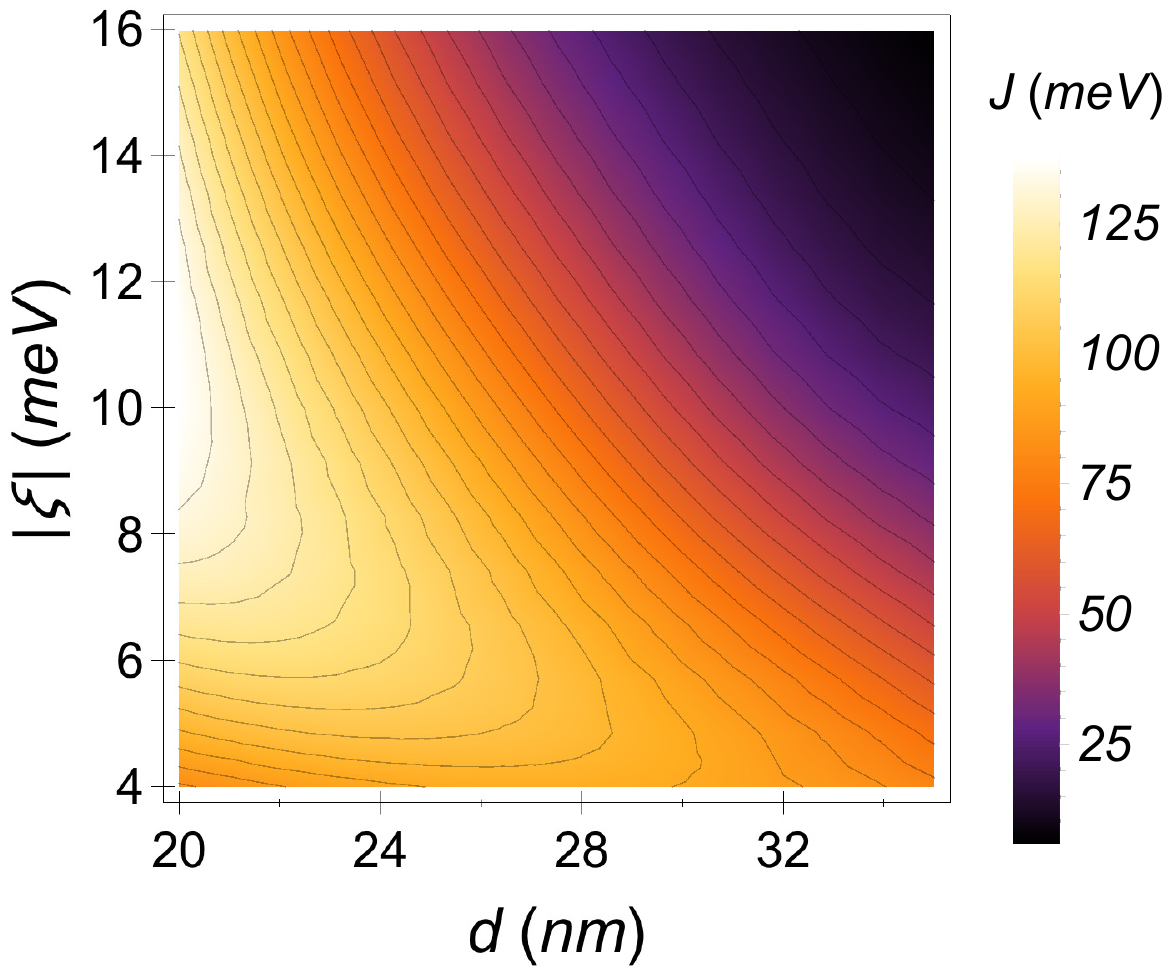}
  \caption{Pseudo-color plot of exchange energy as a function of interdot distance \(d\) (x-axis) and well depth \(|\xi|\) (y-axis). The color scales indicate the strength of the exchange energy \(J\), as indicated by the bar legend on the right.}
  \label{fig:3DPlot}
\end{figure}

\subsection{Exchange coupling as a function of well depth \(\xi\)}
	\label{subsec:depth}
The exchange interaction was calculated as a function of well depth while maintaining constant interdot distance. Fig. \ref{fig:depth} shows that the exchange interaction exhibited the following two features depending on the magnitude of the well depth and interdot distance: (a) the exponential decay of the exchange interaction as a result of the increase of well depth, and (b) the increase in exchange interaction as a result of the increase in the well depth for $|\xi|\lesssim10 meV$. These two features appear at a different range of the well depth depending on the interdot distance. When the well depth was large, the exponentially decaying behavior became the prominent property of the exchange interaction. However, when the well depth was shallow, the increase of $J$ as a function of $|\xi|$ became prominent instead. Therefore, in Fig.~\ref{fig:depth}(b) the exchange interaction exhibited a hill-shaped trend as a function of well depth. The exponential decay of the exchange interaction at a large well depth can be easily explained by the decrease of interdot tunneling and long-range Coulomb interaction due to the stronger confinement of potential wells, which increased with well depth. Nevertheless, the increase of $J$ as a function of $|\xi|$ was dependent on computational details.
Again, the value of the exchange interaction occurred in the range of $meV$, and the reason for this discrepancy has already been discussed in a previous section. Moreover, the calculation also showed that the high orbital levels, in this case $S(2,0)$ and $S(0,2)$, became less relevant for narrow dots (large $\omega_0$) since the difference of $J$'s calculated based on the HL and HM approach was significantly reduced for the narrow dots, which indicated that the exchange interaction was less affected by high orbital levels.
Fig. \ref{fig:depth} also shows that the exchange interaction calculated using the FD states with either CA or bi-quadratic potential overlapped with each other (red and black lines), which indicated that the exchange derived from the FD states was insensitive to the confinement strength exerted by the potential function. However, with regard to the HM calculation, the overlapping started to disappear and those two lines started to split when the well depth was shallower than $10 meV$, which suggests the insensitivity to potential confinement changes as a function of well depth. Thereby, the degree of insensitivity was high for a deep well but started to decrease when the well became shallow. In any case, the exchange interactions predicted by using the FD states were all smaller than those predicted by the CA states.

In Fig.~\ref{fig:3DPlot} we show the exchange interaction calculated by the CA states as a function of both the interdot distance and the well depth in a pseudo-color plot. It is clear that there was a maximum at $|\xi|\approx 10 meV$ and $d\approx 20nm$. The exchange interaction decayed exponentially along both directions, either by increasing $d$ or by increasing $|\xi|$ beyond 10 meV. We note that while this calculation was only performed in one dimension since the exact solution to the two-dimentional Schr\"odinger equation was not available yet. We can expect that the underestimation of the calculations based on the FD states may be even worse in higher dimensions. Therefore, our calculations suggest that the conventional method of using the FD states as the starting point of HL or HM calculations, with two or multi-electron states approximated by the linear combinations of the FD states, should at least be revisited.

\section{Conclusions}
	\label{sec:conclusions}
In this study, the exactly solvable wave functions presented by Caticha \cite{caticha1995construction} were adopted to calculate the exchange interaction $J$ with respect to singlet-triplet qubit $ST_0$ architecture. Compared to the results derived from conventional methods using FD states with a bi-quadratic potential, the exchange interaction $J$ derived from the states of CA, presents several interesting features such as a significant magnitude of $J$ caused by a high interdot probability density and the non-monotonous curve of $J$, which is assumed to arise from the additional shoulder of probability density of the CA state at its decaying tail. These features of exchange coupling were not observed in the $J$ curve calculated by using the FD states; instead, it shows a monotonous trend of exponential decay at a large interdot distance (as a function of interdot distance) and at large depth (as a function of well depth). This phenomenon suggests that some features or magnitude of $J$ may have been overlooked by previous studies on the $ST_0$ qubit or other qubit architectures since only FD states were adopted. One important flaw in adopting FD states was their failure to allow interdot tunneling, which was observed for the Caticha state (Fig. \ref{fig:CaWavefunction}), since it could only solve a single potential well Schr\"odinger equation. Nevertheless, the CA states were limited to one-dimensional wave functions. However, the two-electron states in DQD were well described by two-dimensional wave functions, such as FD states, or sometimes three-dimensional wave functions. Although, it was observed that there were discrepancies between our calculation results and other widely accepted experimental results, such as the $meV$ range of $J$ and the explosion of $J$ when the interdot distance was small, our results have shown that the notion of exactly adopting the solvable Schr\"odinger equation solutions promises a more reasonable calculation for the exchange interaction of the ST$_0$ qubit. However, there is still limited research and solutions to the mathematical equations of the exactly solvable double-well potential, particularly in two- or higher-dimensional scopes. Future work will need to adopt a two-dimensional exactly solvable double-well potential and attempt to solve the exchange coupling $J$ in an analytical manner.

 The work described in this paper is supported by the 
Research Grants Council of the Hong Kong Special Administrative Region, China (No.~CityU 21300116), the National Natural Science Foundation of China (No.~11604277), and the Guangdong Innovative and Entrepreneurial Research Team Program (No.~2016ZT06D348).


\begin{thebibliography}{42}%
\makeatletter
\providecommand \@ifxundefined [1]{%
 \@ifx{#1\undefined}
}%
\providecommand \@ifnum [1]{%
 \ifnum #1\expandafter \@firstoftwo
 \else \expandafter \@secondoftwo
 \fi
}%
\providecommand \@ifx [1]{%
 \ifx #1\expandafter \@firstoftwo
 \else \expandafter \@secondoftwo
 \fi
}%
\providecommand \natexlab [1]{#1}%
\providecommand \enquote  [1]{``#1''}%
\providecommand \bibnamefont  [1]{#1}%
\providecommand \bibfnamefont [1]{#1}%
\providecommand \citenamefont [1]{#1}%
\providecommand \href@noop [0]{\@secondoftwo}%
\providecommand \href [0]{\begingroup \@sanitize@url \@href}%
\providecommand \@href[1]{\@@startlink{#1}\@@href}%
\providecommand \@@href[1]{\endgroup#1\@@endlink}%
\providecommand \@sanitize@url [0]{\catcode `\\12\catcode `\$12\catcode
  `\&12\catcode `\#12\catcode `\^12\catcode `\_12\catcode `\%12\relax}%
\providecommand \@@startlink[1]{}%
\providecommand \@@endlink[0]{}%
\providecommand \url  [0]{\begingroup\@sanitize@url \@url }%
\providecommand \@url [1]{\endgroup\@href {#1}{\urlprefix }}%
\providecommand \urlprefix  [0]{URL }%
\providecommand \Eprint [0]{\href }%
\providecommand \doibase [0]{http://dx.doi.org/}%
\providecommand \selectlanguage [0]{\@gobble}%
\providecommand \bibinfo  [0]{\@secondoftwo}%
\providecommand \bibfield  [0]{\@secondoftwo}%
\providecommand \translation [1]{[#1]}%
\providecommand \BibitemOpen [0]{}%
\providecommand \bibitemStop [0]{}%
\providecommand \bibitemNoStop [0]{.\EOS\space}%
\providecommand \EOS [0]{\spacefactor3000\relax}%
\providecommand \BibitemShut  [1]{\csname bibitem#1\endcsname}%
\let\auto@bib@innerbib\@empty
\bibitem [{\citenamefont {Shor}(1999)}]{shor1999polynomial}%
  \BibitemOpen
  \bibfield  {author} {\bibinfo {author} {\bibfnamefont {P.~W.}\ \bibnamefont
  {Shor}},\ }\href@noop {} {\bibfield  {journal} {\bibinfo  {journal} {SIAM
  review}\ }\textbf {\bibinfo {volume} {41}},\ \bibinfo {pages} {303} (\bibinfo
  {year} {1999})}\BibitemShut {NoStop}%
\bibitem [{\citenamefont {Nielsen}\ and\ \citenamefont
  {Chuang}(2002)}]{nielsen2002quantum}%
  \BibitemOpen
  \bibfield  {author} {\bibinfo {author} {\bibfnamefont {M.~A.}\ \bibnamefont
  {Nielsen}}\ and\ \bibinfo {author} {\bibfnamefont {I.}~\bibnamefont
  {Chuang}},\ }\href@noop {} {\enquote {\bibinfo {title} {Quantum computation
  and quantum information},}\ } (\bibinfo {year} {2002})\BibitemShut {NoStop}%
\bibitem [{\citenamefont {Petta}\ \emph {et~al.}(2005)\citenamefont {Petta},
  \citenamefont {Johnson}, \citenamefont {Taylor}, \citenamefont {Laird},
  \citenamefont {Yacoby}, \citenamefont {Lukin}, \citenamefont {Marcus},
  \citenamefont {Hanson},\ and\ \citenamefont {Gossard}}]{petta2005coherent}%
  \BibitemOpen
  \bibfield  {author} {\bibinfo {author} {\bibfnamefont {J.~R.}\ \bibnamefont
  {Petta}}, \bibinfo {author} {\bibfnamefont {A.~C.}\ \bibnamefont {Johnson}},
  \bibinfo {author} {\bibfnamefont {J.~M.}\ \bibnamefont {Taylor}}, \bibinfo
  {author} {\bibfnamefont {E.~A.}\ \bibnamefont {Laird}}, \bibinfo {author}
  {\bibfnamefont {A.}~\bibnamefont {Yacoby}}, \bibinfo {author} {\bibfnamefont
  {M.~D.}\ \bibnamefont {Lukin}}, \bibinfo {author} {\bibfnamefont {C.~M.}\
  \bibnamefont {Marcus}}, \bibinfo {author} {\bibfnamefont {M.~P.}\
  \bibnamefont {Hanson}}, \ and\ \bibinfo {author} {\bibfnamefont {A.~C.}\
  \bibnamefont {Gossard}},\ }\href@noop {} {\bibfield  {journal} {\bibinfo
  {journal} {Science}\ }\textbf {\bibinfo {volume} {309}},\ \bibinfo {pages}
  {2180} (\bibinfo {year} {2005})}\BibitemShut {NoStop}%
\bibitem [{\citenamefont {Koppens}\ \emph {et~al.}(2008)\citenamefont
  {Koppens}, \citenamefont {Nowack},\ and\ \citenamefont
  {Vandersypen}}]{koppens2008spin}%
  \BibitemOpen
  \bibfield  {author} {\bibinfo {author} {\bibfnamefont {F.}~\bibnamefont
  {Koppens}}, \bibinfo {author} {\bibfnamefont {K.}~\bibnamefont {Nowack}}, \
  and\ \bibinfo {author} {\bibfnamefont {L.}~\bibnamefont {Vandersypen}},\
  }\href@noop {} {\bibfield  {journal} {\bibinfo  {journal} {Phys. Rev. Lett.}\
  }\textbf {\bibinfo {volume} {100}},\ \bibinfo {pages} {236802} (\bibinfo
  {year} {2008})}\BibitemShut {NoStop}%
\bibitem [{\citenamefont {Barthel}\ \emph {et~al.}(2012)\citenamefont
  {Barthel}, \citenamefont {Medford}, \citenamefont {Bluhm}, \citenamefont
  {Yacoby}, \citenamefont {Marcus}, \citenamefont {Hanson},\ and\ \citenamefont
  {Gossard}}]{barthel2012relaxation}%
  \BibitemOpen
  \bibfield  {author} {\bibinfo {author} {\bibfnamefont {C.}~\bibnamefont
  {Barthel}}, \bibinfo {author} {\bibfnamefont {J.}~\bibnamefont {Medford}},
  \bibinfo {author} {\bibfnamefont {H.}~\bibnamefont {Bluhm}}, \bibinfo
  {author} {\bibfnamefont {A.}~\bibnamefont {Yacoby}}, \bibinfo {author}
  {\bibfnamefont {C.~M.}\ \bibnamefont {Marcus}}, \bibinfo {author}
  {\bibfnamefont {M.}~\bibnamefont {Hanson}}, \ and\ \bibinfo {author}
  {\bibfnamefont {A.}~\bibnamefont {Gossard}},\ }\href@noop {} {\bibfield
  {journal} {\bibinfo  {journal} {Phys. Rev. B}\ }\textbf {\bibinfo {volume}
  {85}},\ \bibinfo {pages} {035306} (\bibinfo {year} {2012})}\BibitemShut
  {NoStop}%
\bibitem [{\citenamefont {Maune}\ \emph {et~al.}(2012)\citenamefont {Maune},
  \citenamefont {Borselli}, \citenamefont {Huang}, \citenamefont {Ladd},
  \citenamefont {Deelman}, \citenamefont {Holabird}, \citenamefont {Kiselev},
  \citenamefont {Alvarado-Rodriguez}, \citenamefont {Ross}, \citenamefont
  {Schmitz} \emph {et~al.}}]{maune2012coherent}%
  \BibitemOpen
  \bibfield  {author} {\bibinfo {author} {\bibfnamefont {B.~M.}\ \bibnamefont
  {Maune}}, \bibinfo {author} {\bibfnamefont {M.~G.}\ \bibnamefont {Borselli}},
  \bibinfo {author} {\bibfnamefont {B.}~\bibnamefont {Huang}}, \bibinfo
  {author} {\bibfnamefont {T.~D.}\ \bibnamefont {Ladd}}, \bibinfo {author}
  {\bibfnamefont {P.~W.}\ \bibnamefont {Deelman}}, \bibinfo {author}
  {\bibfnamefont {K.~S.}\ \bibnamefont {Holabird}}, \bibinfo {author}
  {\bibfnamefont {A.~A.}\ \bibnamefont {Kiselev}}, \bibinfo {author}
  {\bibfnamefont {I.}~\bibnamefont {Alvarado-Rodriguez}}, \bibinfo {author}
  {\bibfnamefont {R.~S.}\ \bibnamefont {Ross}}, \bibinfo {author}
  {\bibfnamefont {A.~E.}\ \bibnamefont {Schmitz}},  \emph {et~al.},\
  }\href@noop {} {\bibfield  {journal} {\bibinfo  {journal} {Nature}\ }\textbf
  {\bibinfo {volume} {481}},\ \bibinfo {pages} {344} (\bibinfo {year}
  {2012})}\BibitemShut {NoStop}%
\bibitem [{\citenamefont {Pla}\ \emph {et~al.}(2013)\citenamefont {Pla},
  \citenamefont {Tan}, \citenamefont {Dehollain}, \citenamefont {Lim},
  \citenamefont {Morton}, \citenamefont {Zwanenburg}, \citenamefont {Jamieson},
  \citenamefont {Dzurak},\ and\ \citenamefont {Morello}}]{pla2013high}%
  \BibitemOpen
  \bibfield  {author} {\bibinfo {author} {\bibfnamefont {J.~J.}\ \bibnamefont
  {Pla}}, \bibinfo {author} {\bibfnamefont {K.~Y.}\ \bibnamefont {Tan}},
  \bibinfo {author} {\bibfnamefont {J.~P.}\ \bibnamefont {Dehollain}}, \bibinfo
  {author} {\bibfnamefont {W.~H.}\ \bibnamefont {Lim}}, \bibinfo {author}
  {\bibfnamefont {J.~J.}\ \bibnamefont {Morton}}, \bibinfo {author}
  {\bibfnamefont {F.~A.}\ \bibnamefont {Zwanenburg}}, \bibinfo {author}
  {\bibfnamefont {D.~N.}\ \bibnamefont {Jamieson}}, \bibinfo {author}
  {\bibfnamefont {A.~S.}\ \bibnamefont {Dzurak}}, \ and\ \bibinfo {author}
  {\bibfnamefont {A.}~\bibnamefont {Morello}},\ }\href@noop {} {\bibfield
  {journal} {\bibinfo  {journal} {Nature}\ }\textbf {\bibinfo {volume} {496}},\
  \bibinfo {pages} {334} (\bibinfo {year} {2013})}\BibitemShut {NoStop}%
\bibitem [{\citenamefont {Muhonen}\ \emph {et~al.}(2014)\citenamefont
  {Muhonen}, \citenamefont {Dehollain}, \citenamefont {Laucht}, \citenamefont
  {Hudson}, \citenamefont {Kalra}, \citenamefont {Sekiguchi}, \citenamefont
  {Itoh}, \citenamefont {Jamieson}, \citenamefont {McCallum}, \citenamefont
  {Dzurak} \emph {et~al.}}]{muhonen2014storing}%
  \BibitemOpen
  \bibfield  {author} {\bibinfo {author} {\bibfnamefont {J.~T.}\ \bibnamefont
  {Muhonen}}, \bibinfo {author} {\bibfnamefont {J.~P.}\ \bibnamefont
  {Dehollain}}, \bibinfo {author} {\bibfnamefont {A.}~\bibnamefont {Laucht}},
  \bibinfo {author} {\bibfnamefont {F.~E.}\ \bibnamefont {Hudson}}, \bibinfo
  {author} {\bibfnamefont {R.}~\bibnamefont {Kalra}}, \bibinfo {author}
  {\bibfnamefont {T.}~\bibnamefont {Sekiguchi}}, \bibinfo {author}
  {\bibfnamefont {K.~M.}\ \bibnamefont {Itoh}}, \bibinfo {author}
  {\bibfnamefont {D.~N.}\ \bibnamefont {Jamieson}}, \bibinfo {author}
  {\bibfnamefont {J.~C.}\ \bibnamefont {McCallum}}, \bibinfo {author}
  {\bibfnamefont {A.~S.}\ \bibnamefont {Dzurak}},  \emph {et~al.},\ }\href@noop
  {} {\bibfield  {journal} {\bibinfo  {journal} {Nat. Nanotechnol.}\ }\textbf
  {\bibinfo {volume} {9}},\ \bibinfo {pages} {986} (\bibinfo {year}
  {2014})}\BibitemShut {NoStop}%
\bibitem [{\citenamefont {Kim}\ \emph {et~al.}(2014)\citenamefont {Kim},
  \citenamefont {Shi}, \citenamefont {Simmons}, \citenamefont {Ward},
  \citenamefont {Prance}, \citenamefont {Koh}, \citenamefont {Gamble},
  \citenamefont {Savage}, \citenamefont {Lagally}, \citenamefont {Friesen}
  \emph {et~al.}}]{kim2014quantum}%
  \BibitemOpen
  \bibfield  {author} {\bibinfo {author} {\bibfnamefont {D.}~\bibnamefont
  {Kim}}, \bibinfo {author} {\bibfnamefont {Z.}~\bibnamefont {Shi}}, \bibinfo
  {author} {\bibfnamefont {C.}~\bibnamefont {Simmons}}, \bibinfo {author}
  {\bibfnamefont {D.}~\bibnamefont {Ward}}, \bibinfo {author} {\bibfnamefont
  {J.}~\bibnamefont {Prance}}, \bibinfo {author} {\bibfnamefont {T.~S.}\
  \bibnamefont {Koh}}, \bibinfo {author} {\bibfnamefont {J.~K.}\ \bibnamefont
  {Gamble}}, \bibinfo {author} {\bibfnamefont {D.}~\bibnamefont {Savage}},
  \bibinfo {author} {\bibfnamefont {M.}~\bibnamefont {Lagally}}, \bibinfo
  {author} {\bibfnamefont {M.}~\bibnamefont {Friesen}},  \emph {et~al.},\
  }\href@noop {} {\bibfield  {journal} {\bibinfo  {journal} {Nature}\ }\textbf
  {\bibinfo {volume} {511}},\ \bibinfo {pages} {70} (\bibinfo {year}
  {2014})}\BibitemShut {NoStop}%
\bibitem [{\citenamefont {Qi}\ \emph {et~al.}(2017)\citenamefont {Qi},
  \citenamefont {Wu}, \citenamefont {Ward}, \citenamefont {Prance},
  \citenamefont {Kim}, \citenamefont {Gamble}, \citenamefont {Mohr},
  \citenamefont {Shi}, \citenamefont {Savage}, \citenamefont {Lagally},
  \citenamefont {Eriksson}, \citenamefont {Friesen}, \citenamefont
  {Coppersmith},\ and\ \citenamefont {Vavilov}}]{qi2017effects}%
  \BibitemOpen
  \bibfield  {author} {\bibinfo {author} {\bibfnamefont {Z.}~\bibnamefont
  {Qi}}, \bibinfo {author} {\bibfnamefont {X.}~\bibnamefont {Wu}}, \bibinfo
  {author} {\bibfnamefont {D.~R.}\ \bibnamefont {Ward}}, \bibinfo {author}
  {\bibfnamefont {J.~R.}\ \bibnamefont {Prance}}, \bibinfo {author}
  {\bibfnamefont {D.}~\bibnamefont {Kim}}, \bibinfo {author} {\bibfnamefont
  {J.~K.}\ \bibnamefont {Gamble}}, \bibinfo {author} {\bibfnamefont {R.~T.}\
  \bibnamefont {Mohr}}, \bibinfo {author} {\bibfnamefont {Z.}~\bibnamefont
  {Shi}}, \bibinfo {author} {\bibfnamefont {D.~E.}\ \bibnamefont {Savage}},
  \bibinfo {author} {\bibfnamefont {M.~G.}\ \bibnamefont {Lagally}}, \bibinfo
  {author} {\bibfnamefont {M.~A.}\ \bibnamefont {Eriksson}}, \bibinfo {author}
  {\bibfnamefont {M.}~\bibnamefont {Friesen}}, \bibinfo {author} {\bibfnamefont
  {S.~N.}\ \bibnamefont {Coppersmith}}, \ and\ \bibinfo {author} {\bibfnamefont
  {M.~G.}\ \bibnamefont {Vavilov}},\ }\href@noop {} {\bibfield  {journal}
  {\bibinfo  {journal} {Phys. Rev. B}\ }\textbf {\bibinfo {volume} {96}},\
  \bibinfo {pages} {115305} (\bibinfo {year} {2017})}\BibitemShut {NoStop}%
\bibitem [{\citenamefont {Taylor}\ \emph {et~al.}(2005)\citenamefont {Taylor},
  \citenamefont {Engel}, \citenamefont {D{\"u}r}, \citenamefont {Yacoby},
  \citenamefont {Marcus}, \citenamefont {Zoller},\ and\ \citenamefont
  {Lukin}}]{taylor2005fault}%
  \BibitemOpen
  \bibfield  {author} {\bibinfo {author} {\bibfnamefont {J.}~\bibnamefont
  {Taylor}}, \bibinfo {author} {\bibfnamefont {H.-A.}\ \bibnamefont {Engel}},
  \bibinfo {author} {\bibfnamefont {W.}~\bibnamefont {D{\"u}r}}, \bibinfo
  {author} {\bibfnamefont {A.}~\bibnamefont {Yacoby}}, \bibinfo {author}
  {\bibfnamefont {C.}~\bibnamefont {Marcus}}, \bibinfo {author} {\bibfnamefont
  {P.}~\bibnamefont {Zoller}}, \ and\ \bibinfo {author} {\bibfnamefont
  {M.}~\bibnamefont {Lukin}},\ }\href@noop {} {\bibfield  {journal} {\bibinfo
  {journal} {Nat. Phys.}\ }\textbf {\bibinfo {volume} {1}},\ \bibinfo {pages}
  {177} (\bibinfo {year} {2005})}\BibitemShut {NoStop}%
\bibitem [{\citenamefont {Han}\ \emph {et~al.}(2017)\citenamefont {Han},
  \citenamefont {Chen}, \citenamefont {Cao}, \citenamefont {Li}, \citenamefont
  {Xiao},\ and\ \citenamefont {Guo}}]{han2017radio}%
  \BibitemOpen
  \bibfield  {author} {\bibinfo {author} {\bibfnamefont {T.}~\bibnamefont
  {Han}}, \bibinfo {author} {\bibfnamefont {M.}~\bibnamefont {Chen}}, \bibinfo
  {author} {\bibfnamefont {G.}~\bibnamefont {Cao}}, \bibinfo {author}
  {\bibfnamefont {H.}~\bibnamefont {Li}}, \bibinfo {author} {\bibfnamefont
  {M.}~\bibnamefont {Xiao}}, \ and\ \bibinfo {author} {\bibfnamefont
  {G.}~\bibnamefont {Guo}},\ }\href@noop {} {\bibfield  {journal} {\bibinfo
  {journal} {Sci. China Phys. Mech. Astron.}\ }\textbf {\bibinfo {volume}
  {60}},\ \bibinfo {pages} {057301} (\bibinfo {year} {2017})}\BibitemShut
  {NoStop}%
\bibitem [{\citenamefont {Chen}\ \emph
  {et~al.}(2017{\natexlab{a}})\citenamefont {Chen}, \citenamefont {Wang},
  \citenamefont {Cao}, \citenamefont {Li}, \citenamefont {Xiao},\ and\
  \citenamefont {Guo}}]{chen2017enhanced}%
  \BibitemOpen
  \bibfield  {author} {\bibinfo {author} {\bibfnamefont {B.}~\bibnamefont
  {Chen}}, \bibinfo {author} {\bibfnamefont {B.}~\bibnamefont {Wang}}, \bibinfo
  {author} {\bibfnamefont {G.}~\bibnamefont {Cao}}, \bibinfo {author}
  {\bibfnamefont {H.}~\bibnamefont {Li}}, \bibinfo {author} {\bibfnamefont
  {M.}~\bibnamefont {Xiao}}, \ and\ \bibinfo {author} {\bibfnamefont
  {G.}~\bibnamefont {Guo}},\ }\href@noop {} {\bibfield  {journal} {\bibinfo
  {journal} {Sci. Bull.}\ }\textbf {\bibinfo {volume} {62}},\ \bibinfo {pages}
  {712} (\bibinfo {year} {2017}{\natexlab{a}})}\BibitemShut {NoStop}%
\bibitem [{\citenamefont {Loss}\ and\ \citenamefont
  {DiVincenzo}(1998)}]{loss1998quantum}%
  \BibitemOpen
  \bibfield  {author} {\bibinfo {author} {\bibfnamefont {D.}~\bibnamefont
  {Loss}}\ and\ \bibinfo {author} {\bibfnamefont {D.~P.}\ \bibnamefont
  {DiVincenzo}},\ }\href@noop {} {\bibfield  {journal} {\bibinfo  {journal}
  {Phys. Rev. A}\ }\textbf {\bibinfo {volume} {57}},\ \bibinfo {pages} {120}
  (\bibinfo {year} {1998})}\BibitemShut {NoStop}%
\bibitem [{\citenamefont {DiVincenzo}\ \emph {et~al.}(2000)\citenamefont
  {DiVincenzo}, \citenamefont {Bacon}, \citenamefont {Kempe}, \citenamefont
  {Burkard},\ and\ \citenamefont {Whaley}}]{divincenzo2000universal}%
  \BibitemOpen
  \bibfield  {author} {\bibinfo {author} {\bibfnamefont {D.~P.}\ \bibnamefont
  {DiVincenzo}}, \bibinfo {author} {\bibfnamefont {D.}~\bibnamefont {Bacon}},
  \bibinfo {author} {\bibfnamefont {J.}~\bibnamefont {Kempe}}, \bibinfo
  {author} {\bibfnamefont {G.}~\bibnamefont {Burkard}}, \ and\ \bibinfo
  {author} {\bibfnamefont {K.~B.}\ \bibnamefont {Whaley}},\ }\href@noop {}
  {\bibfield  {journal} {\bibinfo  {journal} {Nature}\ }\textbf {\bibinfo
  {volume} {408}},\ \bibinfo {pages} {339} (\bibinfo {year}
  {2000})}\BibitemShut {NoStop}%
\bibitem [{\citenamefont {Laird}\ \emph {et~al.}(2010)\citenamefont {Laird},
  \citenamefont {Taylor}, \citenamefont {DiVincenzo}, \citenamefont {Marcus},
  \citenamefont {Hanson},\ and\ \citenamefont {Gossard}}]{laird2010coherent}%
  \BibitemOpen
  \bibfield  {author} {\bibinfo {author} {\bibfnamefont {E.~A.}\ \bibnamefont
  {Laird}}, \bibinfo {author} {\bibfnamefont {J.~M.}\ \bibnamefont {Taylor}},
  \bibinfo {author} {\bibfnamefont {D.~P.}\ \bibnamefont {DiVincenzo}},
  \bibinfo {author} {\bibfnamefont {C.~M.}\ \bibnamefont {Marcus}}, \bibinfo
  {author} {\bibfnamefont {M.~P.}\ \bibnamefont {Hanson}}, \ and\ \bibinfo
  {author} {\bibfnamefont {A.~C.}\ \bibnamefont {Gossard}},\ }\href@noop {}
  {\bibfield  {journal} {\bibinfo  {journal} {Phys. Rev. B}\ }\textbf {\bibinfo
  {volume} {82}},\ \bibinfo {pages} {075403} (\bibinfo {year}
  {2010})}\BibitemShut {NoStop}%
\bibitem [{\citenamefont {Medford}\ \emph {et~al.}(2013)\citenamefont
  {Medford}, \citenamefont {Beil}, \citenamefont {Taylor}, \citenamefont
  {Rashba}, \citenamefont {Lu}, \citenamefont {Gossard},\ and\ \citenamefont
  {Marcus}}]{medford2013quantum}%
  \BibitemOpen
  \bibfield  {author} {\bibinfo {author} {\bibfnamefont {J.}~\bibnamefont
  {Medford}}, \bibinfo {author} {\bibfnamefont {J.}~\bibnamefont {Beil}},
  \bibinfo {author} {\bibfnamefont {J.}~\bibnamefont {Taylor}}, \bibinfo
  {author} {\bibfnamefont {E.}~\bibnamefont {Rashba}}, \bibinfo {author}
  {\bibfnamefont {H.}~\bibnamefont {Lu}}, \bibinfo {author} {\bibfnamefont
  {A.}~\bibnamefont {Gossard}}, \ and\ \bibinfo {author} {\bibfnamefont
  {C.~M.}\ \bibnamefont {Marcus}},\ }\href@noop {} {\bibfield  {journal}
  {\bibinfo  {journal} {Phys. Rev. Lett.}\ }\textbf {\bibinfo {volume} {111}},\
  \bibinfo {pages} {050501} (\bibinfo {year} {2013})}\BibitemShut {NoStop}%
\bibitem [{\citenamefont {Shi}\ \emph {et~al.}(2012)\citenamefont {Shi},
  \citenamefont {Simmons}, \citenamefont {Prance}, \citenamefont {Gamble},
  \citenamefont {Koh}, \citenamefont {Shim}, \citenamefont {Hu}, \citenamefont
  {Savage}, \citenamefont {Lagally}, \citenamefont {Eriksson} \emph
  {et~al.}}]{shi2012fast}%
  \BibitemOpen
  \bibfield  {author} {\bibinfo {author} {\bibfnamefont {Z.}~\bibnamefont
  {Shi}}, \bibinfo {author} {\bibfnamefont {C.}~\bibnamefont {Simmons}},
  \bibinfo {author} {\bibfnamefont {J.}~\bibnamefont {Prance}}, \bibinfo
  {author} {\bibfnamefont {J.~K.}\ \bibnamefont {Gamble}}, \bibinfo {author}
  {\bibfnamefont {T.~S.}\ \bibnamefont {Koh}}, \bibinfo {author} {\bibfnamefont
  {Y.-P.}\ \bibnamefont {Shim}}, \bibinfo {author} {\bibfnamefont
  {X.}~\bibnamefont {Hu}}, \bibinfo {author} {\bibfnamefont {D.}~\bibnamefont
  {Savage}}, \bibinfo {author} {\bibfnamefont {M.}~\bibnamefont {Lagally}},
  \bibinfo {author} {\bibfnamefont {M.}~\bibnamefont {Eriksson}},  \emph
  {et~al.},\ }\href@noop {} {\bibfield  {journal} {\bibinfo  {journal} {Phys.
  Rev. Lett.}\ }\textbf {\bibinfo {volume} {108}},\ \bibinfo {pages} {140503}
  (\bibinfo {year} {2012})}\BibitemShut {NoStop}%
\bibitem [{\citenamefont {Cao}\ \emph {et~al.}(2016)\citenamefont {Cao},
  \citenamefont {Li}, \citenamefont {Yu}, \citenamefont {Wang}, \citenamefont
  {Chen}, \citenamefont {Song}, \citenamefont {Xiao}, \citenamefont {Guo},
  \citenamefont {Jiang}, \citenamefont {Hu},\ and\ \citenamefont
  {Guo}}]{PhysRevLett.116.086801}%
  \BibitemOpen
  \bibfield  {author} {\bibinfo {author} {\bibfnamefont {G.}~\bibnamefont
  {Cao}}, \bibinfo {author} {\bibfnamefont {H.-O.}\ \bibnamefont {Li}},
  \bibinfo {author} {\bibfnamefont {G.-D.}\ \bibnamefont {Yu}}, \bibinfo
  {author} {\bibfnamefont {B.-C.}\ \bibnamefont {Wang}}, \bibinfo {author}
  {\bibfnamefont {B.-B.}\ \bibnamefont {Chen}}, \bibinfo {author}
  {\bibfnamefont {X.-X.}\ \bibnamefont {Song}}, \bibinfo {author}
  {\bibfnamefont {M.}~\bibnamefont {Xiao}}, \bibinfo {author} {\bibfnamefont
  {G.-C.}\ \bibnamefont {Guo}}, \bibinfo {author} {\bibfnamefont {H.-W.}\
  \bibnamefont {Jiang}}, \bibinfo {author} {\bibfnamefont {X.}~\bibnamefont
  {Hu}}, \ and\ \bibinfo {author} {\bibfnamefont {G.-P.}\ \bibnamefont {Guo}},\
  }\href@noop {} {\bibfield  {journal} {\bibinfo  {journal} {Phys. Rev. Lett.}\
  }\textbf {\bibinfo {volume} {116}},\ \bibinfo {pages} {086801} (\bibinfo
  {year} {2016})}\BibitemShut {NoStop}%
\bibitem [{\citenamefont {Chen}\ \emph
  {et~al.}(2017{\natexlab{b}})\citenamefont {Chen}, \citenamefont {Wang},
  \citenamefont {Cao}, \citenamefont {Li}, \citenamefont {Xiao}, \citenamefont
  {Guo}, \citenamefont {Jiang}, \citenamefont {Hu},\ and\ \citenamefont
  {Guo}}]{PhysRevB.95.035408}%
  \BibitemOpen
  \bibfield  {author} {\bibinfo {author} {\bibfnamefont {B.-B.}\ \bibnamefont
  {Chen}}, \bibinfo {author} {\bibfnamefont {B.-C.}\ \bibnamefont {Wang}},
  \bibinfo {author} {\bibfnamefont {G.}~\bibnamefont {Cao}}, \bibinfo {author}
  {\bibfnamefont {H.-O.}\ \bibnamefont {Li}}, \bibinfo {author} {\bibfnamefont
  {M.}~\bibnamefont {Xiao}}, \bibinfo {author} {\bibfnamefont {G.-C.}\
  \bibnamefont {Guo}}, \bibinfo {author} {\bibfnamefont {H.-W.}\ \bibnamefont
  {Jiang}}, \bibinfo {author} {\bibfnamefont {X.}~\bibnamefont {Hu}}, \ and\
  \bibinfo {author} {\bibfnamefont {G.-P.}\ \bibnamefont {Guo}},\ }\href@noop
  {} {\bibfield  {journal} {\bibinfo  {journal} {Phys. Rev. B}\ }\textbf
  {\bibinfo {volume} {95}},\ \bibinfo {pages} {035408} (\bibinfo {year}
  {2017}{\natexlab{b}})}\BibitemShut {NoStop}%
\bibitem [{\citenamefont {Burkard}\ \emph {et~al.}(1999)\citenamefont
  {Burkard}, \citenamefont {Loss},\ and\ \citenamefont
  {DiVincenzo}}]{burkard1999coupled}%
  \BibitemOpen
  \bibfield  {author} {\bibinfo {author} {\bibfnamefont {G.}~\bibnamefont
  {Burkard}}, \bibinfo {author} {\bibfnamefont {D.}~\bibnamefont {Loss}}, \
  and\ \bibinfo {author} {\bibfnamefont {D.~P.}\ \bibnamefont {DiVincenzo}},\
  }\href@noop {} {\bibfield  {journal} {\bibinfo  {journal} {Phys. Rev. B}\
  }\textbf {\bibinfo {volume} {59}},\ \bibinfo {pages} {2070} (\bibinfo {year}
  {1999})}\BibitemShut {NoStop}%
\bibitem [{\citenamefont {Hu}\ and\ \citenamefont
  {Sarma}(2000)}]{hu2000hilbert}%
  \BibitemOpen
  \bibfield  {author} {\bibinfo {author} {\bibfnamefont {X.}~\bibnamefont
  {Hu}}\ and\ \bibinfo {author} {\bibfnamefont {S.~D.}\ \bibnamefont {Sarma}},\
  }\href@noop {} {\bibfield  {journal} {\bibinfo  {journal} {Phys. Rev. A}\
  }\textbf {\bibinfo {volume} {61}},\ \bibinfo {pages} {062301} (\bibinfo
  {year} {2000})}\BibitemShut {NoStop}%
\bibitem [{\citenamefont {He}\ \emph {et~al.}(2005)\citenamefont {He},
  \citenamefont {Bester},\ and\ \citenamefont {Zunger}}]{he2005singlet}%
  \BibitemOpen
  \bibfield  {author} {\bibinfo {author} {\bibfnamefont {L.}~\bibnamefont
  {He}}, \bibinfo {author} {\bibfnamefont {G.}~\bibnamefont {Bester}}, \ and\
  \bibinfo {author} {\bibfnamefont {A.}~\bibnamefont {Zunger}},\ }\href@noop {}
  {\bibfield  {journal} {\bibinfo  {journal} {Phys. Rev. B}\ }\textbf {\bibinfo
  {volume} {72}},\ \bibinfo {pages} {195307} (\bibinfo {year}
  {2005})}\BibitemShut {NoStop}%
\bibitem [{\citenamefont {Saraiva}\ \emph {et~al.}(2007)\citenamefont
  {Saraiva}, \citenamefont {Calder{\'o}n},\ and\ \citenamefont
  {Koiller}}]{saraiva2007reliability}%
  \BibitemOpen
  \bibfield  {author} {\bibinfo {author} {\bibfnamefont {A.}~\bibnamefont
  {Saraiva}}, \bibinfo {author} {\bibfnamefont {M.}~\bibnamefont
  {Calder{\'o}n}}, \ and\ \bibinfo {author} {\bibfnamefont {B.}~\bibnamefont
  {Koiller}},\ }\href@noop {} {\bibfield  {journal} {\bibinfo  {journal} {Phys.
  Rev. B}\ }\textbf {\bibinfo {volume} {76}},\ \bibinfo {pages} {233302}
  (\bibinfo {year} {2007})}\BibitemShut {NoStop}%
\bibitem [{\citenamefont {Li}\ \emph {et~al.}(2010)\citenamefont {Li},
  \citenamefont {Cywi{\'n}ski}, \citenamefont {Culcer}, \citenamefont {Hu},\
  and\ \citenamefont {Sarma}}]{li2010exchange}%
  \BibitemOpen
  \bibfield  {author} {\bibinfo {author} {\bibfnamefont {Q.}~\bibnamefont
  {Li}}, \bibinfo {author} {\bibfnamefont {{\L}.}~\bibnamefont {Cywi{\'n}ski}},
  \bibinfo {author} {\bibfnamefont {D.}~\bibnamefont {Culcer}}, \bibinfo
  {author} {\bibfnamefont {X.}~\bibnamefont {Hu}}, \ and\ \bibinfo {author}
  {\bibfnamefont {S.~D.}\ \bibnamefont {Sarma}},\ }\href@noop {} {\bibfield
  {journal} {\bibinfo  {journal} {Phys. Rev. B}\ }\textbf {\bibinfo {volume}
  {81}},\ \bibinfo {pages} {085313} (\bibinfo {year} {2010})}\BibitemShut
  {NoStop}%
\bibitem [{\citenamefont {Yang}\ and\ \citenamefont
  {Sarma}(2011)}]{yang2011low}%
  \BibitemOpen
  \bibfield  {author} {\bibinfo {author} {\bibfnamefont {S.}~\bibnamefont
  {Yang}}\ and\ \bibinfo {author} {\bibfnamefont {S.~D.}\ \bibnamefont
  {Sarma}},\ }\href@noop {} {\bibfield  {journal} {\bibinfo  {journal} {Phys.
  Rev. B}\ }\textbf {\bibinfo {volume} {84}},\ \bibinfo {pages} {121306}
  (\bibinfo {year} {2011})}\BibitemShut {NoStop}%
\bibitem [{\citenamefont {Nielsen}\ \emph {et~al.}(2012)\citenamefont
  {Nielsen}, \citenamefont {Muller},\ and\ \citenamefont
  {Carroll}}]{nielsen2012configuration}%
  \BibitemOpen
  \bibfield  {author} {\bibinfo {author} {\bibfnamefont {E.}~\bibnamefont
  {Nielsen}}, \bibinfo {author} {\bibfnamefont {R.~P.}\ \bibnamefont {Muller}},
  \ and\ \bibinfo {author} {\bibfnamefont {M.~S.}\ \bibnamefont {Carroll}},\
  }\href@noop {} {\bibfield  {journal} {\bibinfo  {journal} {Phys. Rev. B}\
  }\textbf {\bibinfo {volume} {85}},\ \bibinfo {pages} {035319} (\bibinfo
  {year} {2012})}\BibitemShut {NoStop}%
\bibitem [{\citenamefont {Mehl}\ and\ \citenamefont
  {DiVincenzo}(2014)}]{mehl2014inverted}%
  \BibitemOpen
  \bibfield  {author} {\bibinfo {author} {\bibfnamefont {S.}~\bibnamefont
  {Mehl}}\ and\ \bibinfo {author} {\bibfnamefont {D.~P.}\ \bibnamefont
  {DiVincenzo}},\ }\href@noop {} {\bibfield  {journal} {\bibinfo  {journal}
  {Phys. Rev. B}\ }\textbf {\bibinfo {volume} {90}},\ \bibinfo {pages} {195424}
  (\bibinfo {year} {2014})}\BibitemShut {NoStop}%
\bibitem [{\citenamefont {Calderon-Vargas}\ and\ \citenamefont
  {Kestner}(2015)}]{calderon2015directly}%
  \BibitemOpen
  \bibfield  {author} {\bibinfo {author} {\bibfnamefont {F.~A.}\ \bibnamefont
  {Calderon-Vargas}}\ and\ \bibinfo {author} {\bibfnamefont {J.~P.}\
  \bibnamefont {Kestner}},\ }\href@noop {} {\bibfield  {journal} {\bibinfo
  {journal} {Phys. Rev. B}\ }\textbf {\bibinfo {volume} {91}},\ \bibinfo
  {pages} {035301} (\bibinfo {year} {2015})}\BibitemShut {NoStop}%
\bibitem [{\citenamefont {Nielsen}\ \emph {et~al.}(2010)\citenamefont
  {Nielsen}, \citenamefont {Young}, \citenamefont {Muller},\ and\ \citenamefont
  {Carroll}}]{nielsen2010implications}%
  \BibitemOpen
  \bibfield  {author} {\bibinfo {author} {\bibfnamefont {E.}~\bibnamefont
  {Nielsen}}, \bibinfo {author} {\bibfnamefont {R.~W.}\ \bibnamefont {Young}},
  \bibinfo {author} {\bibfnamefont {R.~P.}\ \bibnamefont {Muller}}, \ and\
  \bibinfo {author} {\bibfnamefont {M.}~\bibnamefont {Carroll}},\ }\href@noop
  {} {\bibfield  {journal} {\bibinfo  {journal} {Phys. Rev. B}\ }\textbf
  {\bibinfo {volume} {82}},\ \bibinfo {pages} {075319} (\bibinfo {year}
  {2010})}\BibitemShut {NoStop}%
\bibitem [{\citenamefont {Raith}\ \emph {et~al.}(2011)\citenamefont {Raith},
  \citenamefont {Stano},\ and\ \citenamefont {Fabian}}]{raith2011theory}%
  \BibitemOpen
  \bibfield  {author} {\bibinfo {author} {\bibfnamefont {M.}~\bibnamefont
  {Raith}}, \bibinfo {author} {\bibfnamefont {P.}~\bibnamefont {Stano}}, \ and\
  \bibinfo {author} {\bibfnamefont {J.}~\bibnamefont {Fabian}},\ }\href@noop {}
  {\bibfield  {journal} {\bibinfo  {journal} {Phys. Rev. B}\ }\textbf {\bibinfo
  {volume} {83}},\ \bibinfo {pages} {195318} (\bibinfo {year}
  {2011})}\BibitemShut {NoStop}%
\bibitem [{\citenamefont {Barnes}\ \emph {et~al.}(2011)\citenamefont {Barnes},
  \citenamefont {Kestner}, \citenamefont {Nguyen},\ and\ \citenamefont
  {Sarma}}]{barnes2011screening}%
  \BibitemOpen
  \bibfield  {author} {\bibinfo {author} {\bibfnamefont {E.}~\bibnamefont
  {Barnes}}, \bibinfo {author} {\bibfnamefont {J.}~\bibnamefont {Kestner}},
  \bibinfo {author} {\bibfnamefont {N.}~\bibnamefont {Nguyen}}, \ and\ \bibinfo
  {author} {\bibfnamefont {S.~D.}\ \bibnamefont {Sarma}},\ }\href@noop {}
  {\bibfield  {journal} {\bibinfo  {journal} {Phys. Rev. B}\ }\textbf {\bibinfo
  {volume} {84}},\ \bibinfo {pages} {235309} (\bibinfo {year}
  {2011})}\BibitemShut {NoStop}%
\bibitem [{\citenamefont {Bakker}\ \emph {et~al.}(2015)\citenamefont {Bakker},
  \citenamefont {Mehl}, \citenamefont {Hiltunen}, \citenamefont {Harju},\ and\
  \citenamefont {DiVincenzo}}]{bakker2015validity}%
  \BibitemOpen
  \bibfield  {author} {\bibinfo {author} {\bibfnamefont {M.~A.}\ \bibnamefont
  {Bakker}}, \bibinfo {author} {\bibfnamefont {S.}~\bibnamefont {Mehl}},
  \bibinfo {author} {\bibfnamefont {T.}~\bibnamefont {Hiltunen}}, \bibinfo
  {author} {\bibfnamefont {A.}~\bibnamefont {Harju}}, \ and\ \bibinfo {author}
  {\bibfnamefont {D.~P.}\ \bibnamefont {DiVincenzo}},\ }\href@noop {}
  {\bibfield  {journal} {\bibinfo  {journal} {Phys. Rev. B}\ }\textbf {\bibinfo
  {volume} {91}},\ \bibinfo {pages} {155425} (\bibinfo {year}
  {2015})}\BibitemShut {NoStop}%
\bibitem [{\citenamefont {Culcer}\ and\ \citenamefont
  {Zimmerman}(2013)}]{culcer2013dephasing}%
  \BibitemOpen
  \bibfield  {author} {\bibinfo {author} {\bibfnamefont {D.}~\bibnamefont
  {Culcer}}\ and\ \bibinfo {author} {\bibfnamefont {N.~M.}\ \bibnamefont
  {Zimmerman}},\ }\href@noop {} {\bibfield  {journal} {\bibinfo  {journal}
  {Appl. Phys. Lett.}\ }\textbf {\bibinfo {volume} {102}},\ \bibinfo {pages}
  {232108} (\bibinfo {year} {2013})}\BibitemShut {NoStop}%
\bibitem [{\citenamefont {Jiang}\ \emph {et~al.}(2013)\citenamefont {Jiang},
  \citenamefont {Yang}, \citenamefont {Pan}, \citenamefont {Rossi},
  \citenamefont {Dzurak},\ and\ \citenamefont {Culcer}}]{jiang2013coulomb}%
  \BibitemOpen
  \bibfield  {author} {\bibinfo {author} {\bibfnamefont {L.}~\bibnamefont
  {Jiang}}, \bibinfo {author} {\bibfnamefont {C.}~\bibnamefont {Yang}},
  \bibinfo {author} {\bibfnamefont {Z.}~\bibnamefont {Pan}}, \bibinfo {author}
  {\bibfnamefont {A.}~\bibnamefont {Rossi}}, \bibinfo {author} {\bibfnamefont
  {A.~S.}\ \bibnamefont {Dzurak}}, \ and\ \bibinfo {author} {\bibfnamefont
  {D.}~\bibnamefont {Culcer}},\ }\href@noop {} {\bibfield  {journal} {\bibinfo
  {journal} {Phys. Rev. B}\ }\textbf {\bibinfo {volume} {88}},\ \bibinfo
  {pages} {085311} (\bibinfo {year} {2013})}\BibitemShut {NoStop}%
\bibitem [{\citenamefont {Zimmerman}\ \emph {et~al.}(2017)\citenamefont
  {Zimmerman}, \citenamefont {Huang},\ and\ \citenamefont
  {Culcer}}]{zimmerman2017valley}%
  \BibitemOpen
  \bibfield  {author} {\bibinfo {author} {\bibfnamefont {N.~M.}\ \bibnamefont
  {Zimmerman}}, \bibinfo {author} {\bibfnamefont {P.}~\bibnamefont {Huang}}, \
  and\ \bibinfo {author} {\bibfnamefont {D.}~\bibnamefont {Culcer}},\
  }\href@noop {} {\bibfield  {journal} {\bibinfo  {journal} {Nano Lett.}\
  }\textbf {\bibinfo {volume} {17}},\ \bibinfo {pages} {4461} (\bibinfo {year}
  {2017})}\BibitemShut {NoStop}%
\bibitem [{\citenamefont {Caticha}(1995)}]{caticha1995construction}%
  \BibitemOpen
  \bibfield  {author} {\bibinfo {author} {\bibfnamefont {A.}~\bibnamefont
  {Caticha}},\ }\href@noop {} {\bibfield  {journal} {\bibinfo  {journal} {Phys.
  Rev. A}\ }\textbf {\bibinfo {volume} {51}},\ \bibinfo {pages} {4264}
  (\bibinfo {year} {1995})}\BibitemShut {NoStop}%
\bibitem [{\citenamefont {Chen}\ \emph {et~al.}(2012)\citenamefont {Chen},
  \citenamefont {Wu},\ and\ \citenamefont {Xie}}]{chen2012heun}%
  \BibitemOpen
  \bibfield  {author} {\bibinfo {author} {\bibfnamefont {B.-H.}\ \bibnamefont
  {Chen}}, \bibinfo {author} {\bibfnamefont {Y.}~\bibnamefont {Wu}}, \ and\
  \bibinfo {author} {\bibfnamefont {Q.-T.}\ \bibnamefont {Xie}},\ }\href@noop
  {} {\bibfield  {journal} {\bibinfo  {journal} {J. Phys. A}\ }\textbf
  {\bibinfo {volume} {46}},\ \bibinfo {pages} {035301} (\bibinfo {year}
  {2012})}\BibitemShut {NoStop}%
\bibitem [{\citenamefont {Xie}\ \emph {et~al.}(2015)\citenamefont {Xie},
  \citenamefont {Wang},\ and\ \citenamefont {Fu}}]{xie2015analytical}%
  \BibitemOpen
  \bibfield  {author} {\bibinfo {author} {\bibfnamefont {Q.}~\bibnamefont
  {Xie}}, \bibinfo {author} {\bibfnamefont {L.}~\bibnamefont {Wang}}, \ and\
  \bibinfo {author} {\bibfnamefont {J.}~\bibnamefont {Fu}},\ }\href@noop {}
  {\bibfield  {journal} {\bibinfo  {journal} {Phys. Scripta.}\ }\textbf
  {\bibinfo {volume} {90}},\ \bibinfo {pages} {045204} (\bibinfo {year}
  {2015})}\BibitemShut {NoStop}%
\bibitem [{\citenamefont {Xie}(2012)}]{xie2012new}%
  \BibitemOpen
  \bibfield  {author} {\bibinfo {author} {\bibfnamefont {Q.-T.}\ \bibnamefont
  {Xie}},\ }\href@noop {} {\bibfield  {journal} {\bibinfo  {journal} {J. Phys.
  A}\ }\textbf {\bibinfo {volume} {45}},\ \bibinfo {pages} {175302} (\bibinfo
  {year} {2012})}\BibitemShut {NoStop}%
\bibitem [{\citenamefont {Jelic}\ and\ \citenamefont
  {Marsiglio}(2012)}]{jelic2012double}%
  \BibitemOpen
  \bibfield  {author} {\bibinfo {author} {\bibfnamefont {V.}~\bibnamefont
  {Jelic}}\ and\ \bibinfo {author} {\bibfnamefont {F.}~\bibnamefont
  {Marsiglio}},\ }\href@noop {} {\bibfield  {journal} {\bibinfo  {journal}
  {Eur. J. Phys.}\ }\textbf {\bibinfo {volume} {33}},\ \bibinfo {pages} {1651}
  (\bibinfo {year} {2012})}\BibitemShut {NoStop}%
\bibitem [{\citenamefont {Mu{\~n}oz-Vega}\ \emph {et~al.}(2014)\citenamefont
  {Mu{\~n}oz-Vega}, \citenamefont {L{\'o}pez-Ch{\'a}vez}, \citenamefont
  {Salinas-Hernandez}, \citenamefont {Flores-Godoy},\ and\ \citenamefont
  {Fern{\'a}ndez-Anaya}}]{munoz2014exactly}%
  \BibitemOpen
  \bibfield  {author} {\bibinfo {author} {\bibfnamefont {R.}~\bibnamefont
  {Mu{\~n}oz-Vega}}, \bibinfo {author} {\bibfnamefont {E.}~\bibnamefont
  {L{\'o}pez-Ch{\'a}vez}}, \bibinfo {author} {\bibfnamefont {E.}~\bibnamefont
  {Salinas-Hernandez}}, \bibinfo {author} {\bibfnamefont {J.-J.}\ \bibnamefont
  {Flores-Godoy}}, \ and\ \bibinfo {author} {\bibfnamefont {G.}~\bibnamefont
  {Fern{\'a}ndez-Anaya}},\ }\href@noop {} {\bibfield  {journal} {\bibinfo
  {journal} {Phys. Lett. A}\ }\textbf {\bibinfo {volume} {378}},\ \bibinfo
  {pages} {2070} (\bibinfo {year} {2014})}\BibitemShut {NoStop}%
\end{thebibliography}
\end{document}